\documentclass[12pt,preprint]{aastex}

\shorttitle{Stellar Rotation in Young Clusters II.}
\shortauthors{Huang \& Gies}

\begin{document}

\received{}
\accepted{}

\title{Stellar Rotation in Young Clusters. \\ 
II. Evolution of Stellar Rotation and Surface Helium Abundance}

\author{W. Huang and D. R. Gies\altaffilmark{1}}

\affil{Center for High Angular Resolution Astronomy\\
Department of Physics and Astronomy \\
Georgia State University, P. O. Box 4106, Atlanta, GA  30302-4106;\\
huang@chara.gsu.edu, gies@chara.gsu.edu}

\altaffiltext{1}{Visiting Astronomer, Kitt Peak National Observatory
and Cerro Tololo Interamerican Observatory, 
National Optical Astronomy Observatory, operated by the Association
of Universities for Research in Astronomy, Inc., under contract with
the National Science Foundation.}

\slugcomment{Submitted to ApJ}

\paperid{}


\begin{abstract}
We derive the effective temperatures and gravities of 
461 OB stars in 19 young clusters by fitting 
the H$\gamma$ profile in their spectra.
We use synthetic model profiles for rotating stars to 
develop a method to estimate the polar gravity for these stars, 
which we argue is a useful indicator of their evolutionary status. 
We combine these results with projected rotational velocity
measurements obtained in a previous paper on
these same open clusters.  We find that the more massive
B-stars experience a spin down as predicted by the theories 
for the evolution of rotating stars.  Furthermore, we find 
that the members of binary stars also experience a 
marked spin down with advanced evolutionary state 
due to tidal interactions.  
We also derive non-LTE-corrected helium abundances for most of
the sample by fitting the \ion{He}{1} $\lambda\lambda 4026, 4387, 4471$ lines.
A large number of helium peculiar stars are found among cooler stars with
$T_{\rm eff} < 23000$~K.  The analysis of the high mass stars 
($8.5 M_\odot < M < 16 M_\odot$) shows that
the helium enrichment process progresses through the 
main sequence (MS) phase and is greater among the 
faster rotators.  This discovery
supports the theoretical claim that rotationally induced 
internal mixing is the main cause of surface chemical anomalies that
appear during the MS phase.  The lower mass stars appear 
to have slower rotation rates among the low gravity objects, 
and they have a large proportion of helium peculiar stars. 
We suggest that both properties are due to their youth.
The low gravity stars are probably pre-main sequence objects 
that will spin up as they contract.  These young objects 
very likely host a remnant magnetic field from their natal cloud, 
and these strong fields sculpt out surface regions with unusual 
chemical abundances.
\end{abstract}



\setcounter{footnote}{1}

\section{Introduction}                              

The rotational properties of the early-type OB stars are of 
key importance in both observational and theoretical studies of 
massive stars, because this group has the largest range of 
measured rotational velocities and because their evolutionary 
tracks in the Hertzsprung-Russell diagram (HRD) are closely 
linked to rotation.   Theoretical studies by \citet{heg00} and \citet{mey00} 
demonstrate that the evolutionary paths of rapidly rotating 
massive stars can be very different from those for non-rotating stars. 
Rapid rotation can trigger strong mixing inside massive stars, 
extend the core hydrogen-burning lifetime, 
significantly increase the luminosity, and change the chemical composition on
the stellar surface with time.  These evolutionary models predict that 
single (non-magnetic) stars will spin down during the main sequence (MS) phase
due to angular momentum loss in the stellar wind, a net increase in the moment   
of inertia, and an increase in stellar radius. 

However, the direct comparison of the observational data with theoretical 
predictions is difficult because rotation causes the stellar flux to 
become dependent on orientation with respect to the spin axis. 
Rotation changes a star in two fundamental ways: the photospheric shape 
is altered by the centrifugal force \citep{col63} and the local 
effective temperature drops from pole to equator resulting in a 
flux reduction near the equator called gravity darkening \citep{von24}. 
Consequently, the physical parameters of temperature and gravity (and perhaps 
microturbulent velocity) become functions of the colatitude angle from the pole. 
The brightness and color of the star will then depend on the
orientation of its spin axis to the observer \citep*{col77,col91}.  
Evidence of gravity darkening is found in observational studies of eclipsing binary systems
\citep{cla98,cla03} and, more recently, in the direct angular resolution of the 
rapid rotating star $\alpha$~Leo (B7~V) with the CHARA Array optical  
long-baseline interferometer \citep{mca05}. 

This is the second paper in which we investigate the 
rotational properties of B-type stars in a sample of 
19 young open clusters.  In our first paper (\citealt{hua06} = Paper~I), 
we presented estimates of the projected rotational velocities
of 496 stars obtained by comparing observed and synthetic \ion{He}{1}
and \ion{Mg}{2} line profiles.  We discussed evidence of 
changes in the rotational velocity distribution with cluster age, 
but because of the wide range in stellar mass within the samples for 
each cluster it was difficult to study evolutionary effects due
to the mixture of unevolved and evolved stars present.  
Here we present an analysis of the stellar properties of each 
individual star in our sample that we can use to search for 
the effects of rotation among the entire collection of cluster stars. 
We first describe in \S2 a procedure to derive the stellar effective 
temperature $T_{\rm eff}$ and surface gravity $\log g$, 
and then we present in \S3 a method to estimate the polar 
gravity of rotating stars, which we argue is a very good indicator of 
their evolutionary state.  With these key stellar parameters determined, 
we discuss the evolution of stellar rotation and surface helium abundance
in \S4 and \S5, respectively.  We conclude with a summary of our findings 
in \S6. 

 
\section{Effective Temperature and Gravity from H$\gamma$}       

The stellar spectra of B-stars contain many line features that are sensitive 
to temperature $T_{\rm eff}$ and gravity $\log g$.  The hydrogen Balmer 
lines in the optical are particularly useful diagnostic features throughout 
the B-type and late O-type spectral classes.  \citet{gie92}, for example, 
developed an iterative scheme using fits of the H$\gamma$ profile 
(with pressure sensitive wings) and a reddening-free Str\"{o}mgren 
photometric index (mainly sensitive to temperature) to determine 
both the stellar gravity and effective temperature. 
Unfortunately, the available Str\"{o}mgren photometric data for our sample 
are far from complete and have a wide range in accuracy.  
Thus, we decided to develop a procedure to derive both temperature
and gravity based solely upon a fit of the H$\gamma$ line profile
that is recorded in all our spectra (see Paper~I).
We show in Figure~1 how the predicted equivalent width of H$\gamma$ varies 
as a function of temperature and gravity.  The line increases in strength 
with decreasing temperature (to a maximum in the A-stars), and for a given
temperature it also increases in strength with increasing gravity due 
to the increase in the pressure broadening (linear Stark effect) of the 
line wings.   This dependence on gravity is shown in Figure~2 where we 
plot a series of model profiles that all have the same equivalent width 
but differ significantly in line width.  Thus, the combination of line
strength and width essentially offers two parameters that lead 
to a unique model fit dependent on specific values of temperature and gravity.  
This method of matching the Balmer line profiles to derive both parameters 
was presented by \citet{leo97} in their study of the chemical 
abundances in three He-weak stars, and tests suggest that H$\gamma$ fitting 
provides reliable parameter estimates for stars in the temperature range 
from 10000 to 30000~K.  The only prerequisite for the 
application of this method is an accurate estimate of the star's projected 
rotational velocity, $V \sin i$, which we have already obtained in Paper~I. 
The H$\gamma$ profile is the best single choice among the Balmer sequence 
for our purpose, because it is much less affected by incipient emission
that often appears in the H$\alpha$ and H$\beta$ lines among the Be stars
(often rapid rotators) and it is better isolated from other transitions 
in the series compared to the higher order Balmer lines so that only 
one hydrogen line needs to be considered in our synthetic spectra.

\begin{figure}
\plotone{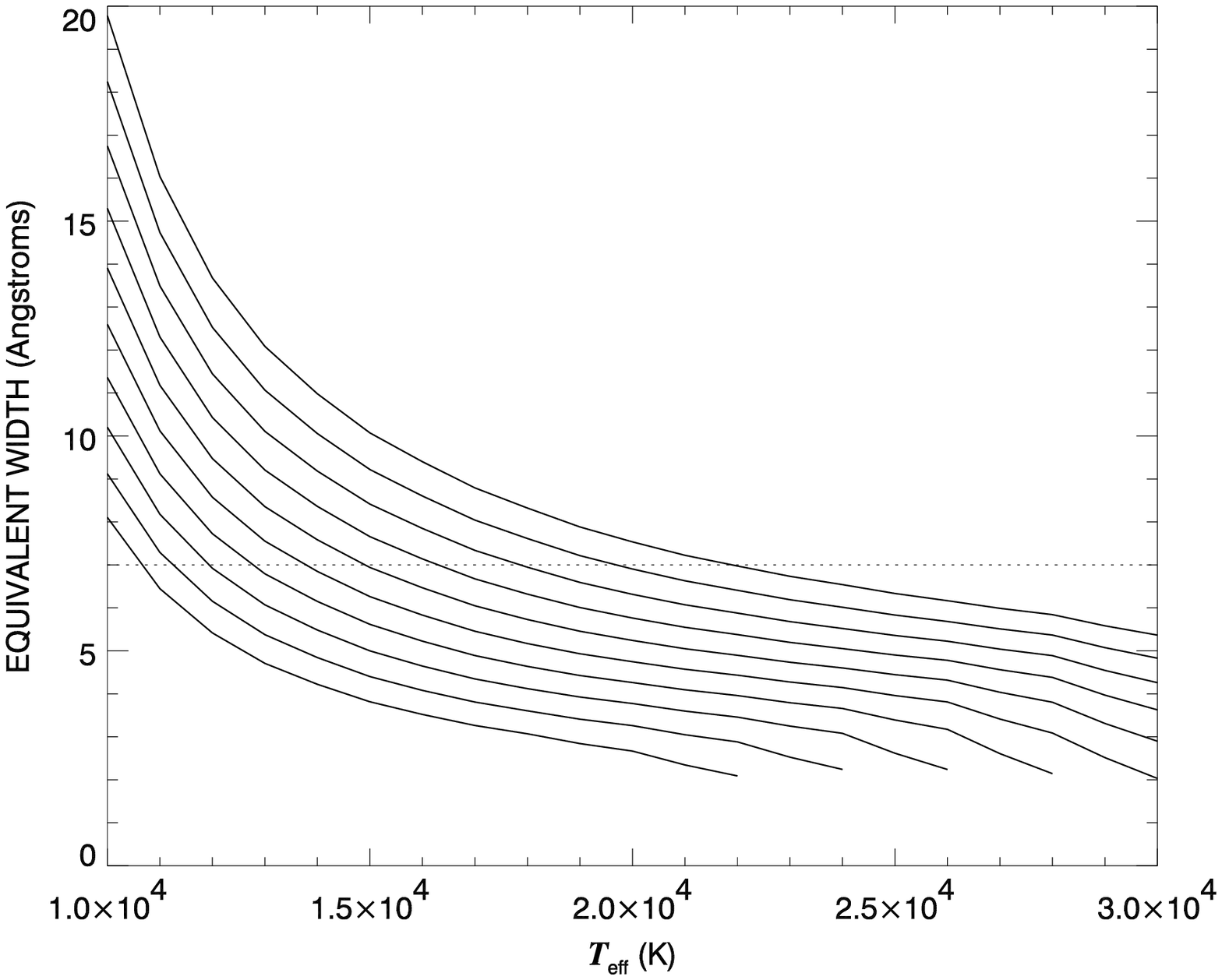}
\caption{
The equivalent width of the H$\gamma$ line (including all strong metallic
lines in the vicinity) plotted as a function of $T_{\rm eff}$.
Each line corresponds to a specific value of $\log g$
that ranges from $\log g = 2.6$ ({\it bottom}) to
$\log g = 4.4$ ({\it top}) in increments of 0.2 dex.
The H$\gamma$ profiles for the intersection points of
the curves and the dotted line ($W_\lambda = 7$ \AA) are
plotted in Figure~\ref{fig2}.
}
\label{fig1}
\end{figure}

\placefigure{fig1}     

\begin{figure}
\plotone{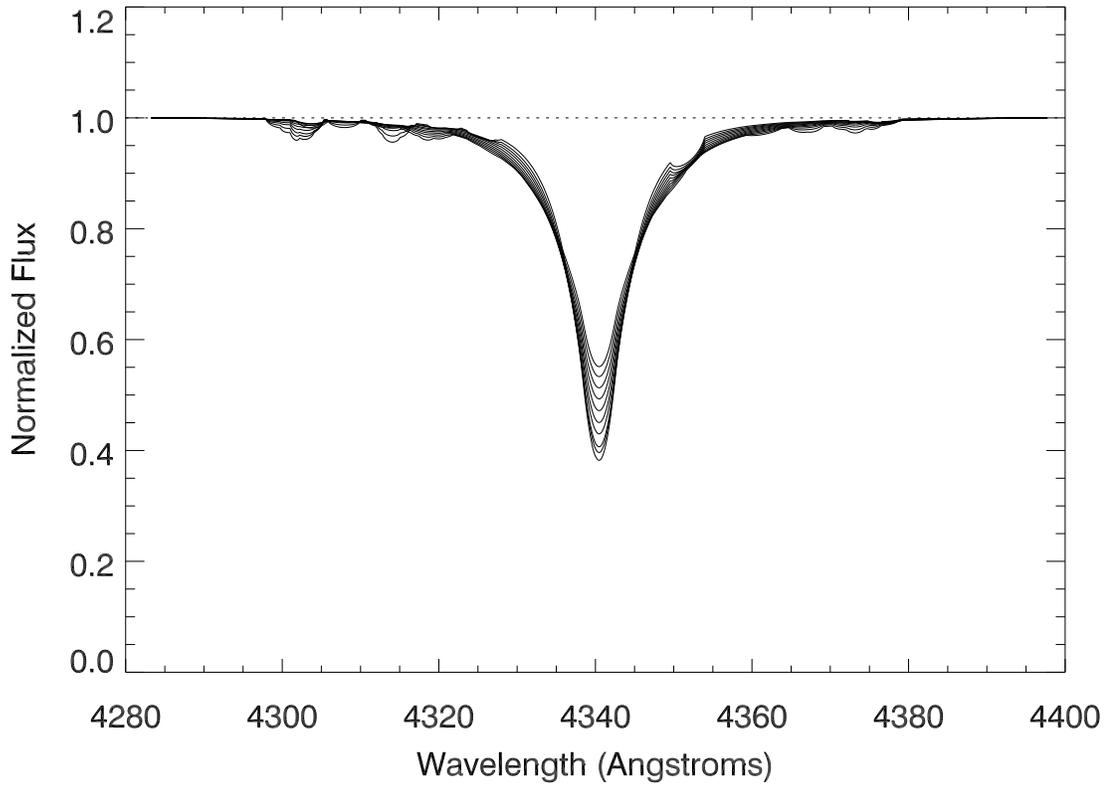}
\caption{
A series of model H$\gamma$ profiles
($V \sin i = 150$ km~s$^{-1}$) with constant equivalent
width (7 \AA) for a locus of $(T_{\rm eff}, \log g)$ points
(defined in Fig.~\ref{fig1}, $T_{\rm eff} =$ 10666, 11255,
11937, 12778, 13786, 14908, 16300, 17865, 19696, 21897 K
for $\log g =$ 2.6, 2.8, $\ldots$, 4.4, respectively).
The profiles vary from narrow at low gravity
to broad-winged at high gravity due to
collisional broadening (linear Stark effect).
} 
\label{fig2}
\end{figure}

\placefigure{fig2}     

The synthetic H$\gamma$ spectra were calculated using a grid of line 
blanketed LTE model atmospheres derived from the ATLAS9 program written 
by Robert Kurucz.  These models assume solar abundances and a microturbulent 
velocity of 2 km~s$^{-1}$, and they were made over a grid of effective 
temperature from $T_{\rm eff}=10000$~K to 30000~K at intervals of 2000~K and over 
a grid of surface gravity from $\log g = 2.6$ to 4.4 at increments of 0.2 dex.  
Then a series of specific intensity spectra were calculated 
using the code SYNSPEC \citep{hub95} for each of these models for a 
cosine of the angle between the surface normal and
line of sight between 0.05 and 1.00 at steps of 0.05.
Finally we used our surface integration code for a rotating star 
(usually with 40000 discrete surface area elements; see Paper~I) 
to calculate the flux spectra from the 
visible hemisphere of the star, and these spectra were convolved 
with the appropriate instrumental broadening function for direct 
comparison with the observed profiles. 

One key issue that needs to be considered in this procedure is 
the line blending in the H$\gamma$ region.  Our sample includes 
stars with effective temperatures ranging from 10000~K to 
30000~K, and the spectral region surrounding H$\gamma$ 
for such stars will include other metallic lines from 
species such as \ion{Ti}{2}, \ion{Fe}{1}, \ion{Fe}{2}, and \ion{Mg}{2}
(in spectra of stars cooler than 12000~K) and from species 
like \ion{O}{2}, \ion{S}{3}, and \ion{Si}{3} (in spectra of
stars hotter than 22000~K).  Many of these lines will be completely 
blended with H$\gamma$, particularly among the spectra of the 
rapid rotators whose metallic profiles will be shallow and broad.  
Neglecting these blends would lead to the introduction of 
systemic errors in the estimates of temperature and gravity
(at least in some temperature domains).   We consulted the 
Web library of B-type synthetic spectra produced by Gummersbach and 
Kaufer\footnote{http://www.lsw.uni-heidelberg.de/cgi-bin/websynspec.cgi} 
\citep{gum98}, and we included in our input line list all the transitions 
shown in their spectra that attain an equivalent width $> 30$ m\AA~ 
in the H$\gamma$ region ($4300 - 4380$\AA).  We assumed a 
solar abundance for these metal lines in each case, and any 
profile errors introduced by deviations from solar abundances
for these weak lines in the actual spectra will be small in comparison to
those errors associated with the observational noise and with 
$V \sin i$ measurement.

We begin by considering the predicted H$\gamma$ profiles for a 
so-called ``virtual star'' by which we mean a star having a spherical shape, 
a projected equatorial rotational velocity equal to a given $V \sin i$, and 
constant gravity and temperature everywhere on its surface. 
Obviously, the concept of a ``virtual star'' does not 
correspond to reality (see \S3), but the application 
of this concept allows us to describe a star using only three parameters: 
$T_{\rm eff}$, $\log g$, and $V \sin i$.  These fitting parameters 
will correspond to some hemispheric average in the real case, and 
can therefore be used as starting points for detailed analysis. 
The procedure for the derivation of temperature and gravity begins 
by assigning a $V \sin i$ to the virtual star model, and then 
comparing the observed H$\gamma$ line profile with a set of synthesized, 
rotationally broadened profiles for the entire temperature-gravity grid.
In practice, we start by measuring the equivalent width of the 
observed H$\gamma$ feature, and then construct a series of interpolated  
temperature-gravity models with this H$\gamma$ equivalent width and 
a range in line broadening (see Fig.~2).  We find the $\chi^2$ 
minimum in the residuals between the observed and model profile sequence, 
and then make a detailed search in the vicinity of this minimum for the 
best fit values of temperature and gravity that correspond to the global 
minimum of the $\chi^2$ statistic. 

We were able to obtain estimates of $T_{\rm eff}$ and $\log g$ for 
461 stars from the sample of cluster stars described in Paper~I. 
Our results are listed in Table~1, which is available in complete 
form in the electronic edition of this paper.  
The columns here correspond to: 
(1) cluster name;
(2) WEBDA index number \citep{mer03} 
(for those stars without a WEBDA index number, we assign them
the same large number ($> 10000$) as we did in Paper~I);
(3) $T_{\rm eff}$; 
(4) error in $T_{\rm eff}$;
(5) $\log g$;
(6) error in $\log g$;
(7) $V\sin i$ (from Paper~I);
(8) estimated polar gravity $\log g_{\rm polar}$ (\S3);
(9 -- 11) log of the He abundance relative to the solar value
as derived from \ion{He}{1} $\lambda\lambda 4026, 4387, 4471$, 
respectively (\S5); 
(12) mean He abundance; and 
(13) inter-line standard deviation of the He abundance. 
Examples of the final line profile fits for three 
stars are shown in Figure~3.   Their corresponding contour plots 
of the residuals from the fit $\chi^2$  
are plotted in Figure~4, where we clearly see that our temperature - 
gravity fitting scheme leads to unambiguous parameter estimates 
and errors.   Each contour interval represents 
an increase in the residuals from the best fit as the ratio 
\begin{displaymath}
{\chi^2_\nu - \chi^2_{\nu ~{\rm min}}} \over
{\chi^2_{\nu ~{\rm min}}/N}
\end{displaymath}
where $N$ represents the number of wavelength points used 
in making the fit ($N\approx 180$), and the specific contours plotted
from close to far from each minimum correspond to ratio values of 
1, 3, 5, 10, 20, 40, 80, 160, 320, 640, and 1280.  
The contours reflect only the errors introduced by the observed 
noise in the spectrum, but we must also account for the 
propagation of errors in the $T_{\rm eff}$ and $\log g$ estimates
due to errors in our $V\sin i$ estimate from Paper~I. 
The average error for $V \sin i$ is about 10~km~s$^{-1}$, so we artificially
increased and decreased the $V \sin i$ measurements by
this value and used the same procedure to derive new temperature and
gravity estimates and hence the propagated errors in these quantities. 
We found that the $V \sin i$ related errors depend on $T_{\rm eff}$, 
$\log g$, and $V \sin i$, and the mean values are 
about $\pm 200$~K for $T_{\rm eff}$ and $\pm 0.02$ for $\log g$.
Our final estimated errors for temperature and $\log g$ (Table~1) 
are based upon the quadratic sum of the $V\sin i$ propagated 
errors and the errors due to intrinsic noise in the observed 
spectrum.  We emphasize that these errors, given in columns 4 and 6 of Table 1, 
represent only the formal errors of the fitting procedure, and they do not
account for possible systematic error sources, such as those 
related to uncertainties in the continuum rectification fits, 
distortions in the profiles caused by disk or wind emission, 
and limitations of the models (static atmospheres with a uniform 
application of microturbulence).   

\placetable{tab1}      

\begin{figure}
\plotone{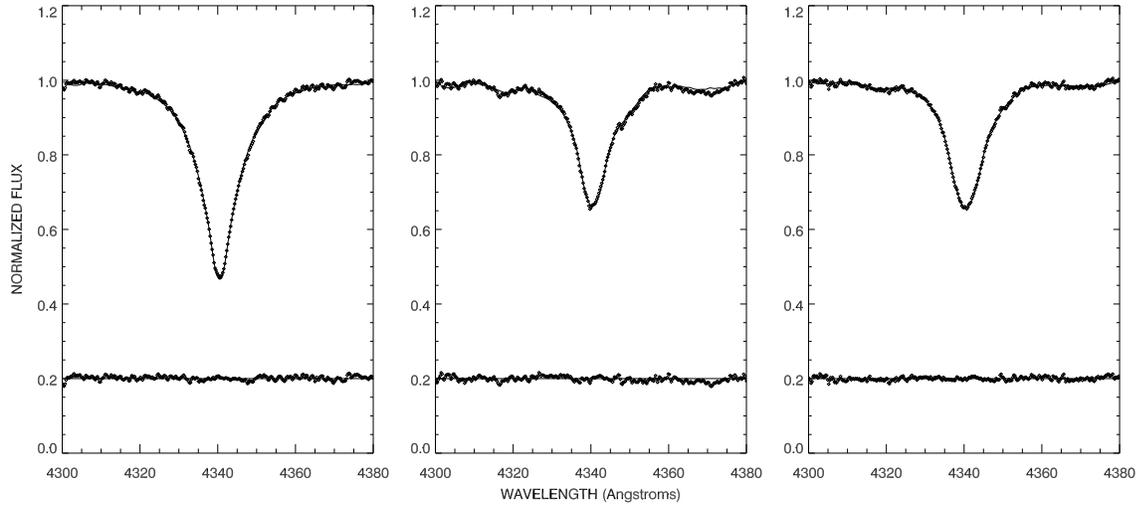}
\caption{
The H$\gamma$ line fitting results for the stars 
({\it from left to right}) 
NGC~1502 \#23 ($T_{\rm eff} = 15286\pm113$~K, $\log g = 3.951\pm0.019$),
NGC~869 \#803 ($T_{\rm eff} = 24838\pm354$~K, $\log g = 3.891\pm0.033$), and
NGC~884 \#2255 ($T_{\rm eff} = 20360\pm178$~K, $\log g = 3.560\pm0.022$).  
The thin solid lines show the fits while the 
plus signs indicate the observed profiles.  The residuals 
from the fit are shown at the bottom of each plot
(offset to 0.2 for clarity).} 
\label{fig3}
\end{figure}

\placefigure{fig3}     

\begin{figure}
\plotone{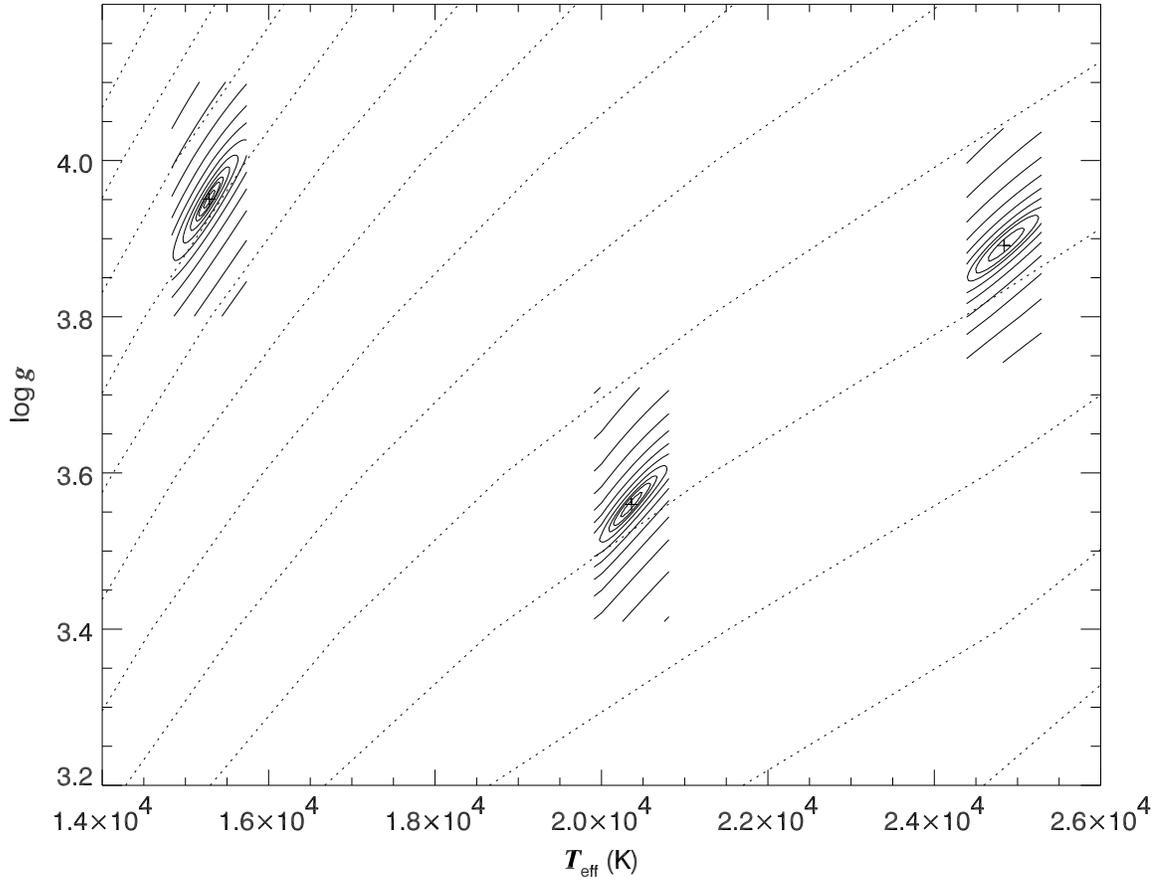}
\caption{
$\chi^2$ residual contour plots for the same stars shown 
in Fig.~3 ({\it from left to right}: NGC~1502 \#23, 
NGC~884 \#2255, and NGC~869 \#803).  The solid lines show 
contours of increasing error residuals relative to the 
minimum value (at the center of the contours), while the
dotted lines are loci of constant equivalent width (see Fig.~1).}
\label{fig4}
\end{figure}

\placefigure{fig4}     

There are several ways in which we can check on the results of 
this H$\gamma$ fitting method, for example, by directly comparing 
the temperatures and gravities with those derived in earlier 
studies, by comparing the derived temperatures with the 
dereddened $(B-V)$ colors, and by comparing the temperatures 
with observed spectral classifications.  Unfortunately, only 
a small number of our targets have prior estimates of 
temperature and gravity.  The best set of common targets 
consists of nine stars in NGC~3293 and NGC~4755 that were 
studied by \citet{mat02}, who estimated the stellar temperatures 
and gravities using Str\"{o}mgren photometry (where the Balmer 
line pressure broadening is measured by comparing fluxes in 
broad and narrowband H$\beta$ filters).  The mean temperature 
and gravity differences for these nine stars are 
$<(T_{\rm eff}(HG)-T_{\rm eff}(M)) / T_{\rm eff}(HG)> = 0.4 \pm 6.5 \%$ and 
$<\log g (HG) - \log g (M) > = 0.00 \pm 0.15$ dex. 
Thus, the parameters derived from the H$\gamma$ method appear
to agree well with those based upon Str\"{o}mgren photometry.

We obtained $(B-V)$ color index data for 441 stars in our sample 
from the WEBDA database. The sources of the photometric data for
each cluster are summarized in order of increasing cluster 
age in Table~2.  Then the intrinsic color
index $(B-V)_0$ for each star was calculated using the mean reddening
$E(B-V)$ of each cluster \citep*{lok01}.  All 441 stars are plotted in Figure~5 according
to their H$\gamma$-derived temperatures and derived color indices.
An empirical relationship between star's surface temperature and its 
intrinsic color is also plotted as a solid line in Figure~5.  This relationship
is based upon the average temperatures for B spectral
subtypes (of luminosity classes IV and V) from \citet{und79} and the intrinsic
colors for these spectral subtypes from \citet{fit70}.  Most of the
stars are clustered around the empirical line, which indicates that our
derived temperatures are consistent with the photometric data.
However, a small number of stars (less than 10\%) in Figure~5 
have colors that are significantly different 
from those expected for cluster member stars.
There are several possible explanations for these stars: (1) they are non-member
stars, either foreground (in the far left region of Fig.~5 due to
over-correction for reddening) or background (in the far right region due to
under-correction for reddening); (2) the reddening distribution of some
clusters may be patchy, so the applied average $E(B-V)$ may over- or
underestimate the reddening of some member stars; or (3) the stars 
may be unresolved binaries with companions of differing color.

\placetable{tab2}      

\begin{figure}
\plotone{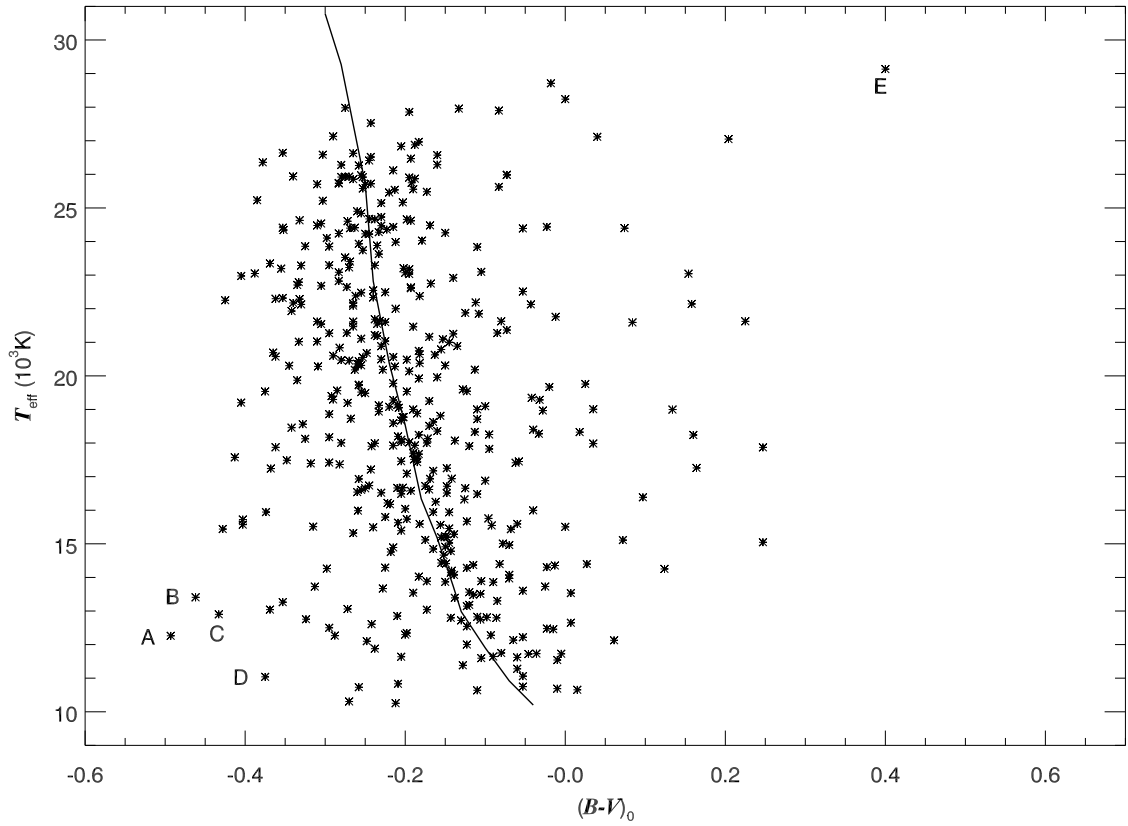}
\caption{
A plot of $T_{\rm eff}$ versus $(B-V)_0$ for 441 sample stars.  
The empirical relationship between stellar effective 
temperature and intrinsic color is indicated by the solid line \citep{fit70,und79}.
Those stars marked by letters are NGC~2244 \#1254 (A), 
IC~1805 \#1433 (B), NGC~2244 \#88 (C), NGC~869 \#748 (D), 
and IC~2944 \#105 (E).
}
\label{fig5}
\end{figure}

\placefigure{fig5}     

We also obtained the MK spectral subtypes of 162 stars in our sample
from the ``MK selected'' category of the WEBDA database. In Figure~6,
most of stars appear to follow the empirical relationship between
spectral subtype and effective temperature found by \citet{boh81},
though the scatter gets larger for hotter stars.  We checked the spectra
of the most discrepant stars (marked by letters) in the figure, and
found that the spectral subtypes of stars C through H were definitely
misclassified.  However, the spectra of Tr~16 \#23 (A) and IC~1805 \#118
(B) show the \ion{He}{2} $\lambda 4200$ feature, which appears only 
in O-star spectra. \citet{aue72} demonstrated that model hydrogen
Balmer line profiles made with non-LTE considerations become significantly different
from those calculated assuming LTE for $T_{\rm eff} > 30000$~K 
(the upper boundary of our temperature grid). 
The equivalent width of the non-LTE profiles decreases with 
increasing temperature much more slowly than 
that of the LTE profiles. Therefore, our H-gamma line fitting method 
based on LTE model atmospheres will lead to an underestimate of the temperature (and 
the gravity) when fits are made of O-star spectra.  However, our sample includes
only a small number of hot O-stars, and, thus, the failure to derive a reliable
surface temperature and gravity for them will not impact significantly our
statistical analysis below.  We identify the 22 problematical O-star cases
(that we found by inspection for the presence of \ion{He}{2} $\lambda 4200$)
by an asterisk in column~1 of Table~1.  These O-stars are omitted in the 
spin and helium abundance discussions below. 

\begin{figure}
\plotone{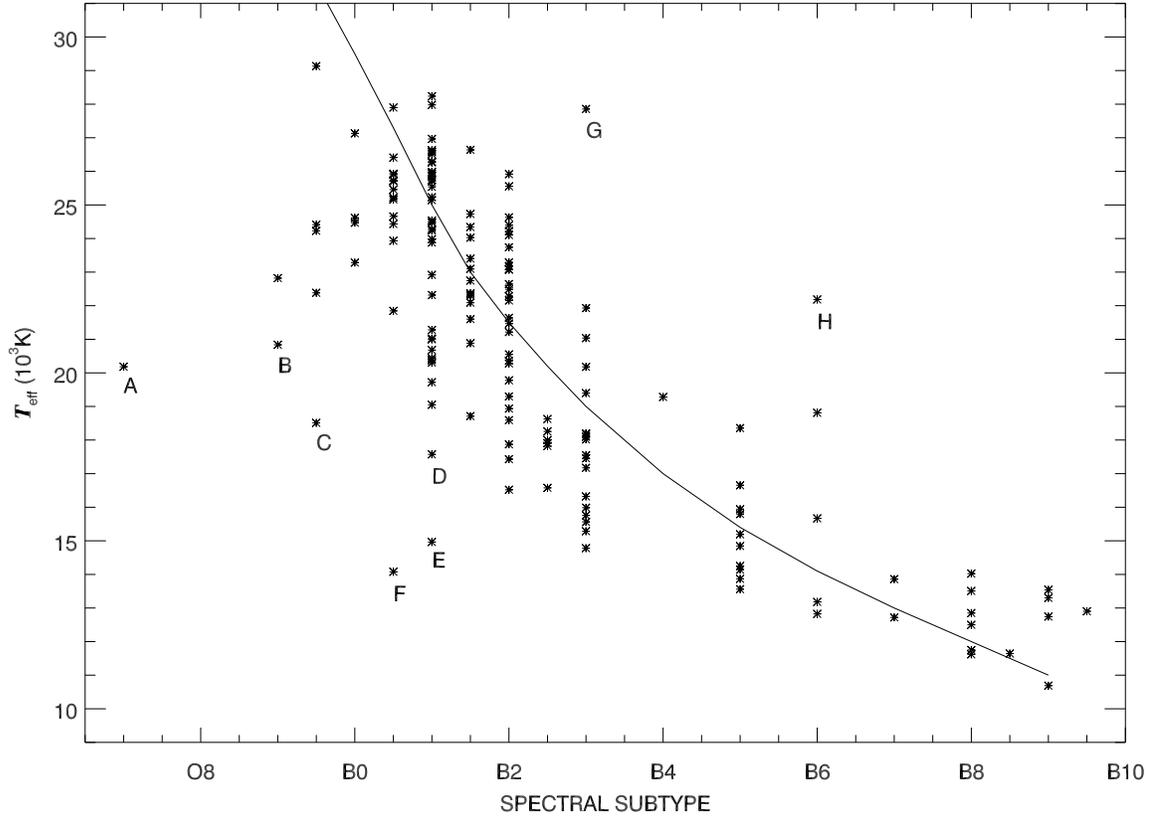}
\caption{
A plot of $T_{\rm eff}$ versus spectral subtype for 162 sample stars.
The relationship between spectral subtype and effective temperature
for OB stars from \citet{boh81} is plotted as a solid line.
Those stars marked by letters are Tr~16 \#23 (A),
IC~1805 \#118 (B), IC~2944 \#48 (C), Tr~16 \#20 (D),
IC~2944 \#110 (E), IC~2944 \#102 (F), NGC~869 \#847 (G), and
NGC~457 \#128 (H).
}
\label{fig6}
\end{figure}

\placefigure{fig6}     


\section{Polar Gravity of Rotating Stars}           

The surface temperature and gravity of a
rotating star vary as functions of the polar colatitude
because of the shape changes due to centrifugal forces
and the associated gravity darkening.  Thus, the estimates 
of temperature and gravity we obtained from the H$\gamma$ 
profile (\S2) represent an average of these parameters
over the visible hemisphere of a given star.  Several questions
need to be answered before we can use these derived values for further
analysis: 
(1) What is the meaning of the stellar temperature and gravity 
derived from the H$\gamma$ fitting method for the case of a rotating star? 
(2) What is the  relationship between our 
derived temperature/gravity values and the distribution of temperature/gravity
on the surface of a real rotating star?  In other words, what kind of average will 
the derived values represent? 
(3) Can we determine the evolutionary
status of rotating stars from the derived temperatures and
gravities as we do in the analysis of non-rotating stars?

In order to answer the first two questions, we would need to 
apply our temperature/gravity determination method to some
real stars whose properties, such as the surface distribution of
temperature and gravity, the orientation of the spin axis in space, and
the projected equatorial velocity, are known to us. However, 
with the exception of $\alpha$~Leo \citep{mca05}, 
we have no OB rotating stars with such reliable data that we can use to
test our method. The alternative is to model the
rotating stars, and then apply our method to their model spectra.

The key parameters to describe a model of a rotating star include
the polar temperature, stellar mass, polar radius, inclination angle, and
$V \sin i$ (see the example of our study of $\alpha$~Leo; \citealt{mca05}).
The surface of the model star is then
calculated based on Roche geometry (i.e., assuming that the 
interior mass is concentrated like a point source) and the 
surface temperature distribution is determined by the 
von Zeipel theorem ($T\propto g^{1/4}_{\rm eff}$; \citealt{von24}). 
Thus, we can use our surface integration code
(which accounts for limb darkening, gravity darkening, and rotational 
changes in the projected area and orientation of the surface elements)  
to synthesize the H$\gamma$ line profile for a given 
inclination angle and $V \sin i$, and we can compare the 
temperature and gravity estimates from our ``virtual star''
approach with the actual run of temperature and gravity on 
the surface of the model rotating star. 

The physical parameters of the nine models we have chosen for this test 
are listed in Table~3.  They are representative of high-mass (models 1, 2, 3),
mid-mass (models 4, 5, 6), and low-mass (models 7, 8, 9) stars in our sample
at different evolutionary stages between the zero age main sequence (ZAMS) 
and terminal age main sequence (TAMS).  The evolutionary stage is closely 
related to the value of the polar gravity $\log g_{\rm polar}$ (see below). 
Table~3 gives the model number, stellar mass, polar radius, polar 
temperature, polar gravity, and critical velocity (at which centripetal 
and gravitation accelerations are equal at the equator).  
We examined the predicted H$\gamma$ profiles for each model 
assuming a range of different combinations of inclination angle and $V \sin i$. 
Theoretical studies of the interiors of rotating stars 
show that the polar radius of a rotating star depends only weakly on angular
velocity (at least for the case of uniform rotation) and usually is 
$<3\%$ different from its value in the case of a non-rotating star
\citep*{sac70,jac05}.  Thus, we assume a constant polar radius for each model. 
We show part of our test results (only for model \#1) in Table~4 
where we list various temperature and gravity estimates for 
a range in assumed inclination angle (between the spin axis and line of sight) 
and projected rotational velocity.    The $T_{\rm eff}$ and 
and $\log g$ derived from fits of the model H$\gamma$ profile 
(our ``virtual star'' approach outlined in \S2) are given in columns 
3 and 4, respectively, and labeled by the subscript {\it msr}. 
These are compared with two kinds of averages of physical values
made by integrations over the visible hemisphere of the model star.  
The first set is for a geometrical mean given by
\begin{displaymath}
<x> = \int x \hat{r} \cdot d\vec{s} \Bigg/ \int \hat{r} \cdot d\vec{s}
\end{displaymath}
where $x$ represents either $T$ or $\log g$ and the integral is 
over the projected area elements given by the dot product of the 
unit line of sight vector $\hat{r}$ and the area element surface 
normal vector $\vec{s}$.   These geometrically defined averages 
are given in columns 5 and 6 and denoted by a subscript {\it geo}.
The next set corresponds to a flux weighted mean given by 
\begin{displaymath}
<x> = \int x I_\lambda \hat{r} \cdot d\vec{s} \Bigg/ \int I_\lambda\hat{r} \cdot d\vec{s}
\end{displaymath}
where $I_\lambda$ is the monochromatic specific intensity from the area element, 
and these averages are listed in columns 7 and 8 with the 
subscript {\it flux}.  Finally we provide an average model temperature~\citep{mey97}
that is independent of inclination and based on the stellar luminosity
\begin{displaymath}
<T_L> = (\int T^4 ds \Bigg/ \int ds)^{1/4}
\end{displaymath}
that is given in column 9.  The final column 10 gives the difference
between the model polar gravity and the measured average gravity, 
$\delta \log g = \log g_{\rm polar} - \log g_{\rm msr}$.  
There is reasonably good agreement between the temperature and
gravity estimates from our ``virtual star'' H$\gamma$ fit
measurements and those from the different model averages, which  
provides some assurance that our method does yield meaningful 
measurements of the temperatures and gravities of rotating stars. 
The listings in Table~4 show the expected trend that as the 
rotation speed increases, the equatorial regions become more 
extended and cooler, resulting in lower overall temperatures and 
gravities.  These effects are largest at an inclination of $90^\circ$ 
where the equatorial zone presents the largest projected area. 

\placetable{tab3}      

\placetable{tab4}      


We can estimate reliably the evolutionary status of a 
non-rotating star by plotting its position in a color-magnitude 
diagram or in its spectroscopic counterpart of a temperature-gravity diagram. 
However, the introduction of rotation makes many of these 
observed quantities dependent on the inclination of the spin axis
\citep{col91} so that position in the HRD is no longer uniquely 
related to a star of specific mass, age, and rotation. 
Furthermore, theoretical models suggest that very rapid rotators might
have dramatically different evolutionary paths than those of 
non-rotating stars \citep{heg00,mey00}, and for some mass ranges  
there are no available theoretical predictions at all for the evolutionary 
tracks of rapid rotators.  Without a reliable observational parameter 
for stellar evolutionary status, it is very difficult to investigate 
the evolution of rotating stars systematically.

The one parameter of a rotating star that is not greatly affected by its rotation is
the polar gravity.  During the entire MS phase, the change of
polar gravity for a rotating star can be attributed to evolutionary effects
almost exclusively.  For example, models of non-rotating stars \citep{sch92}
indicate that the surface gravity varies from $\log g = 4.3$ at the ZAMS 
to $\log g = 3.5$ at the TAMS for a mass range from 2 to 15 $M_\odot$, 
i.e., for the majority of MS B-type stars, and similar results are found 
for the available rotating stellar models \citep{heg00,mey00}. 
Thus, the polar gravity is a good indicator of the evolutionary state 
and it is almost independent of stellar mass among the B-stars.
Rotating stars with different masses but similar polar gravity 
can be treated as a group with a common evolutionary status.  This grouping 
can dramatically increase the significance of statistical results related to  
stellar evolutionary effects when the size of a sample is limited.

We can use the model results given above to help estimate the 
polar gravity for each of the stars in our survey. 
Our measured quantities are $V \sin i$ (Paper~I) and 
$T_{\rm eff}$ and $\log g$ as derived from the H$\gamma$ line fit. 
It is clear from the model results in Table~4 that the 
measured $\log g$ values for a given model will generally 
be lower than the actual polar gravity (see final column in Table~4)
by an amount that depends on $V \sin i$ and inclination angle.  
Unfortunately we cannot derive the true value of the polar gravity for an individual
star from the available data without knowing its spin inclination angle.
Thus, we need to find an alternative way to estimate the polar 
gravity within acceptable errors.  The last column of Table~4 
shows that the difference $\delta \log g = \log g_{\rm polar} - \log g_{\rm msr}$ 
for a specific value of $V\sin i$ changes slowly with inclination angle 
until the angle is so low that the equatorial velocity $(V\sin i) /\sin i$
approaches the critical rotation speed (corresponding to an 
equatorially extended star with a mean gravity significantly 
lower than the polar value).   This suggests that we can average the 
corrections $\delta \log g$ over all possible inclination angles for 
a model at a given $V \sin i$, and then just apply this mean 
correction to our results on individual stars with the same $V \sin i$ value 
to obtain their polar gravity.  As shown in Table~4, this simplification 
of ignoring the specific inclination of stars to estimate their $\log g_{\rm polar}$ values 
will lead to small errors in most cases ($< 0.03$ dex).  The exceptional cases
are those for model stars with equatorial velocities close to the critical value,  
and such situations are generally rare in our sample.

We gathered the model results for $T_{msr}$, $\log g_{msr}$, and 
$\delta \log g$ as a function of inclination $i$ for each model (Table~3) 
and each grid value of $V\sin i$.  We then formed averages of each 
of these three quantities by calculating weighted means over the 
grid values of inclination.  The integrating weight includes two
factors: (1) the factor $\propto \sin i$ to account for the 
probability of the random distribution of spin axes in space; 
(2) the associated probability for the frequency of the implied 
equatorial velocity among our sample of B-stars. Under these
considerations, the mean of a variable $x$ with a specific value
of $V \sin i$ would be
\begin{displaymath}
<x>\big|_{_{V \sin i}} = \frac{\int_{i_{\rm min}}^{\pi/2} 
x|_{_{V \sin i}} P_v(\frac{V \sin i}{\sin i}) \cot i\, di}
{\int_{i_{\rm min}}^{\pi/2} P_v(\frac{V \sin i}{\sin i}) \cot i\, di}
\end{displaymath}
where $P_v$ is the equatorial velocity probability distribution of our sample, 
deconvolved from the $V \sin i$ distribution (see Paper~I), 
and $i_{\rm min}$ is the minimum inclination that corresponds to 
critical rotation at the equator.  Our final 
inclination-averaged means are listed in Table~5 for each model 
and $V\sin i$ pair.   We applied these corrections to each
star in the sample by interpolating in each of these models 
to the observed value of $V\sin i$ and then by making a 
bilinear interpolation in the resulting $V\sin i$ specific 
pairs of $(T_{msr}, \log g_{msr})$ to find the appropriate 
correction term $\delta \log g$ needed to estimate the 
polar gravity.  The resulting polar gravities are listed 
in column~8 of Table~1.

\placetable{tab5}      


\section{Evolution of Stellar Rotation}             

Theoretical models \citep{heg00,mey00} indicate that single rotating stars 
experience a long-term spin down during their MS phase due to angular 
momentum loss by stellar wind and a net increase of the moment of inertia. 
The spin down rate is generally larger in the more massive
stars and those born with faster rotational velocities. 
A spin down may also occur in close binaries due to tidal forces 
acting to bring the stars into synchronous rotation \citep*{abt02}. 
On the other hand, these models also predict that a rapid increase of
rotation velocity can occur at the TAMS caused by an overall contraction 
of the stellar core.  In some cases where the wind mass loss rate is low, 
this increase may bring stars close to the critical velocity. 

Here we examine the changes in the rotational velocity distribution 
with evolution by considering how these distributions vary with polar gravity.  
Since our primary goal is to compare the observed distributions with 
the predictions about stellar rotation evolution for single stars, 
we need to restrict the numbers of binary systems in our working sample.  
We began by excluding all stars that have double-line features in their spectra, 
since these systems have neither reliable $V \sin i$ measurements nor 
reliable temperature and gravity estimates.  
We then divided the rest of our sample into two groups,
single stars (325 objects) and single-lined binaries (78 objects, identified
using the same criterion adopted in the Paper~I, $\Delta V_r > 30$ km~s$^{-1}$).
(Note that we omitted stars from the clusters 
Tr~14 and Tr~16 because we have only single-night observations for these two  
and we cannot determine which stars are spectroscopic binaries.  We
also omitted the O-stars mentioned in \S2 because of uncertainties 
in their derived temperatures and gravities.)  
The stars in these two groups are plotted in the $\log T_{\rm eff} - \log g_{\rm polar}$
plane in Figure~7 (using asterisks for single stars and triangles for binaries).
We also show a set of evolutionary tracks for non-rotating stellar
models with masses from 2.5 $M_\odot$ to 15 $M_\odot$ \citep{sch92} as
indicators of evolutionary status.  The current published data on similar 
evolutionary tracks for rotating stars are restricted to the high mass end 
of this diagram.  However, since the differences between the evolutionary 
tracks for rotating and non-rotating models are modest except for those cases close  
to critical rotation, the use of non-rotating stellar evolutionary tracks should be adequate 
for the statistical analysis that follows.  Figure~7 shows that most 
of the sample stars are located between the ZAMS ($\log g_{\rm polar} = 4.3\pm0.1$) 
and the TAMS ($\log g_{\rm polar} = 3.5\pm0.1$), and the low mass stars appear to be less evolved.
This is the kind of distribution that we would expect for stars
selected from young Galactic clusters.  There are few targets with unusually 
large $\log g_{\rm polar}$ that may be double-lined spectroscopic binaries observed at 
times when line blending makes the H$\gamma$ profile appear very wide.  
For example, the star with the largest gravity ($\log g_{\rm polar}= 4.68$) is 
NGC~2362 \#10008 (= GSC 0654103398), and this target is a radial velocity variable 
and possible binary \citep{hua06}. 

\begin{figure}
\plotone{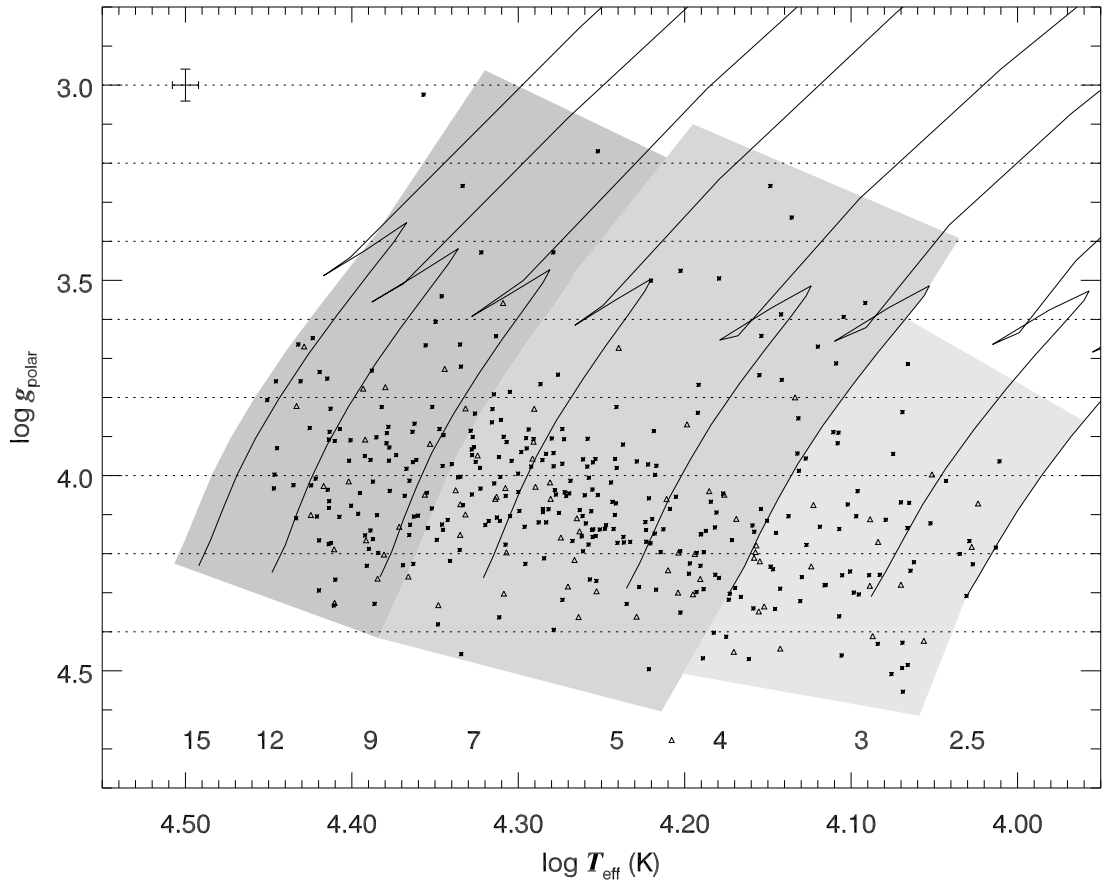}
\caption{
The distribution of the stars in the $\log T_{\rm eff} - \log g_{\rm polar}$ 
plane.  Asterisks are assigned to single stars ($N = 325$) while
triangles are used for binary systems ($N = 78$).  The average errors
in $\log T_{\rm eff}$ and $\log g_{\rm polar}$ are indicated in the upper-left corner.
The solid lines are the evolutionary tracks for non-rotating stellar 
models \citep{sch92}, marked by the mass ($M_\odot$) at the bottom.
The three shaded areas outline the three mass groups corresponding to
the three panels shown in Figure 10.}
\label{fig7}
\end{figure}

\placefigure{fig7}     

We show a similar plot for all rapid rotators in our sample ($V \sin i > 200$ km~s$^{-1}$)
in Figure~8.  These rapid rotators are almost all concentrated in
a band close to the ZAMS.  This immediately suggests that stars form as rapid
rotators and spin down through the MS phase as predicted by the theoretical  
models.  Note that there are three rapid rotators found near the TAMS 
(from cool to hot, the stars are NGC~7160 \#940, NGC~2422 \#125, and
NGC~457 \#128; the latter two are Be stars), and a 
few more such outliers appear in TAMS region if we lower the boundary 
on the rapid rotator group to $V \sin i > 180$ km~s$^{-1}$.  
Why are these three stars separated from all the other rapid rotators? 
One possibility is that they were born as
extremely rapid rotators, so they still have a relatively large amount of
angular momentum at the TAMS compared to other stars.  However, this argument
cannot explain why there is such a clear gap in Figure~8 between these few 
evolved rapid rotators and the large number of young rapid rotators. 
Perhaps these stars are examples of those experiencing a spin up 
during the core contraction that happens near the TAMS.   The scarcity  
of such objects is consistent with the predicted short duration of the spin up phase. 
They may also be examples of stars spun up by mass transfer in close binaries 
(Paper~I).

\begin{figure}
\plotone{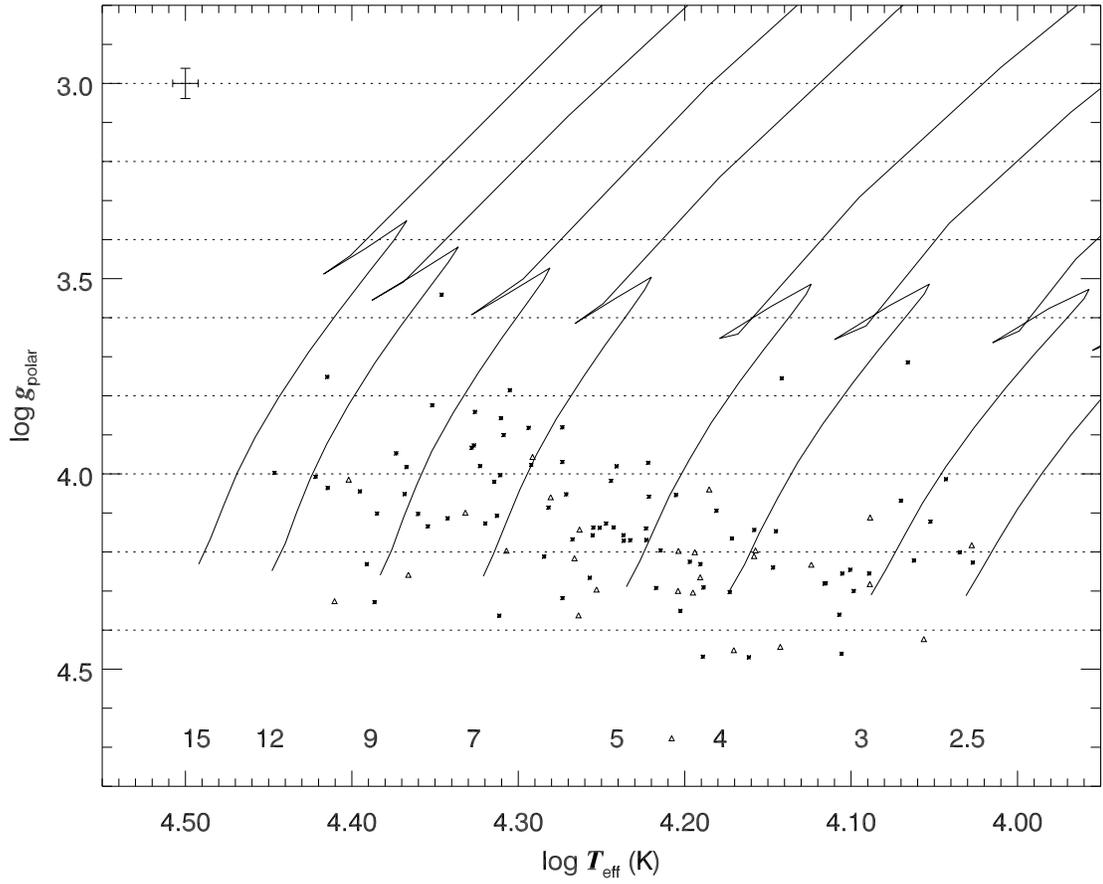}
\caption{
The distribution of the stars with $V \sin i > 200$ km~s$^{-1}$
in the $\log T_{\rm eff} - \log g_{\rm polar}$ plane in the same
format as Fig.~7. }
\label{fig8}
\end{figure}

\placefigure{fig8} 

We next consider the statistics of the rotational velocity distribution 
as a function of evolutionary state by plotting 
diagrams of $V \sin i$ versus $\log g_{\rm polar}$.
Figure~9 shows the distribution of the single stars in our sample in
the $V \sin i - \log g_{\rm polar}$ plane.  These stars were grouped into
0.2 dex bins of $\log g_{\rm polar}$, and the mean and the range within 
one standard deviation of the mean for each bin are plotted as a solid 
line and gray-shaded zone, respectively.  The mean 
$V \sin i$ decreases from $193\pm14$ km~s$^{-1}$ near the ZAMS to 
$88\pm24$ km~s$^{-1}$ near the TAMS. 
If we assume that the ratio of $<V> / <V \sin i> = 4/\pi$  
holds for our sample, then the mean equatorial
velocity for B type stars is $246\pm18$ km~s$^{-1}$ at ZAMS and 
$112\pm31$ km~s$^{-1}$ at TAMS. 
This subsample of single stars was further divided into 
three mass categories (shown by the three shaded areas
in Fig.~7): the high mass group (88 stars, $8.5 M_\odot < M \leq 
16 M_\odot$) is shown in the top panel of Figure~10; the middle mass group 
(174 stars, $4 M_\odot < M \leq 8.5 M_\odot$) in the middle panel, 
and the low mass group
(62 stars, $2.5 M_\odot < M < 4 M_\odot$) in the bottom panel.
All three groups show a spin down trend with decreasing
polar gravity, but their slopes differ. 
The high mass group has a shallow spin down beginning 
from a relatively small initial mean of $V \sin i = 137\pm48$ 
km~s$^{-1}$ (or $<V>=174$ km~s$^{-1}$). The two bins 
($\log g = 3.5, 3.7$, total 15 stars) around the TAMS still have a relatively high
mean $V \sin i$ value, $106\pm29$ km~s$^{-1}$ (or $<V>=134$ km~s$^{-1}$). 
This group has an average mass of 11~$M_\odot$, and it is the 
only mass range covered by current theoretical studies of rotating stars.
The theoretical calculations \citep{heg00,mey00} of the spin
down rate agrees with our statistical results for the high mass group.
\citet{heg00} show (in their Fig.~10) that a star of mass $12 M_\odot$ 
starting with $V = 205$ km~s$^{-1}$ will spin down to 160 km~s$^{-1}$
at the TAMS. \citet{mey00} find a similar result for the same mass model:
for $V = 200$ km~s$^{-1}$ at ZAMS, the equatorial velocity declines to 141 km~s$^{-1}$ at TAMS.  

\begin{figure}
\plotone{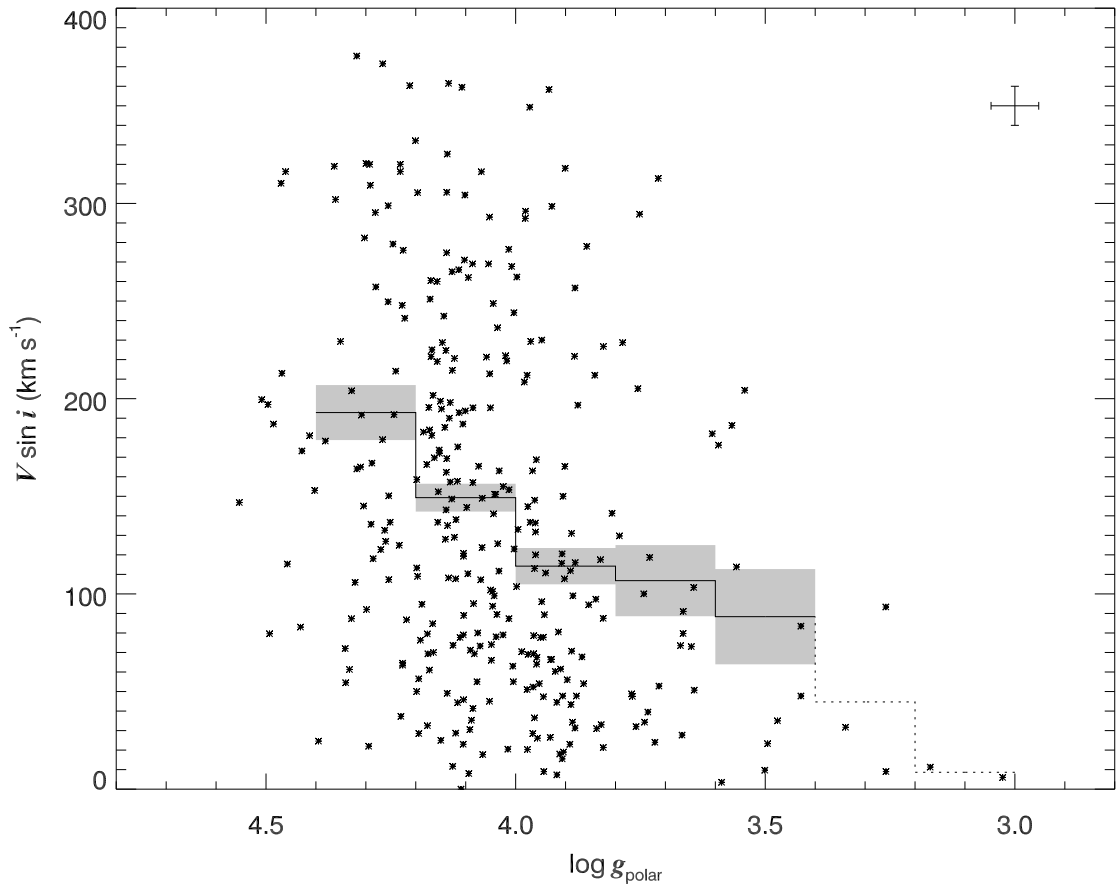}
\caption{
The distribution of the single stars in our sample
in the $(\log g_{\rm polar}, V \sin i)$ plane.  The average errors in
$V \sin i$ and $\log g_{\rm polar}$ are plotted in the upper-right corner.
The solid line shows the mean $V \sin i$ of each 0.2 dex bin of 
$\log g_{\rm polar}$ that contains 6 or more measurements while 
the dotted line shows the same for the rest of bins. 
The shaded areas enclose the associated error of the mean 
in each bin.
}
\label{fig9}
\end{figure}

\placefigure{fig9} 

\begin{figure}
\epsscale{0.68}
\plotone{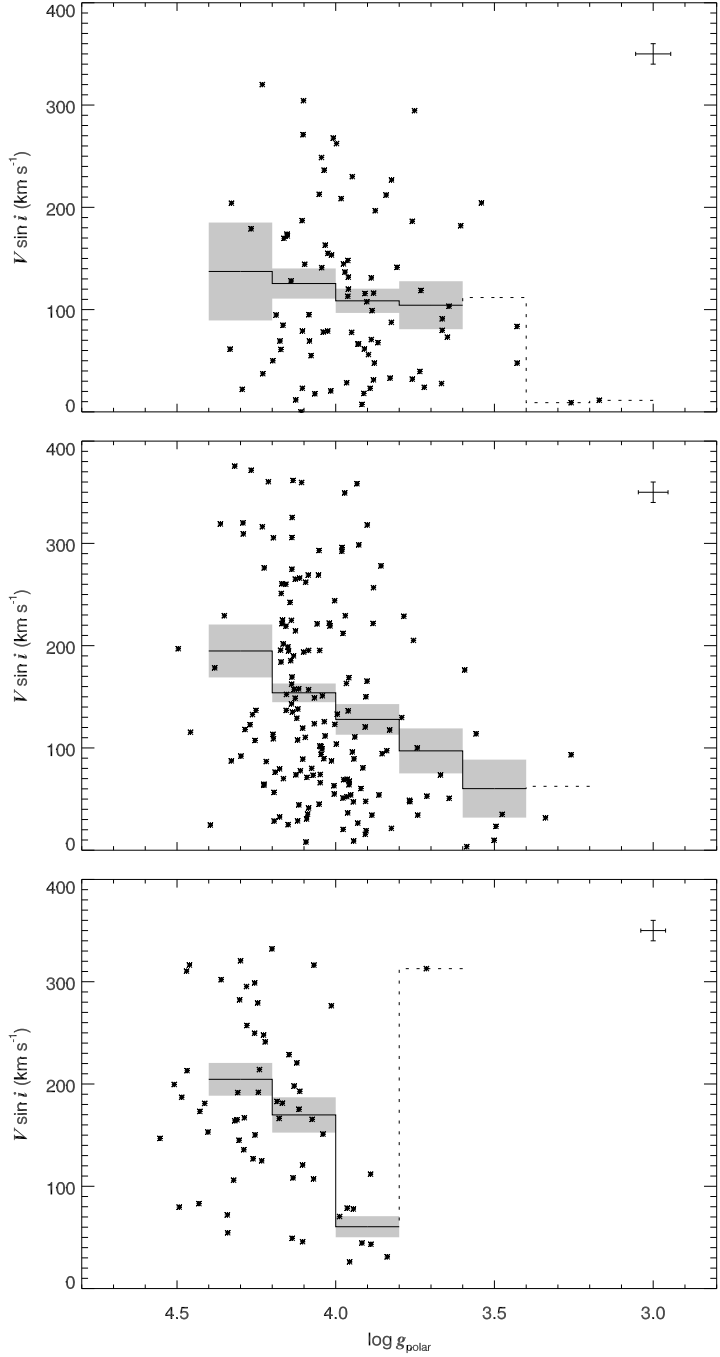}
\caption{
The same $(\log g_{\rm polar}, V \sin i)$ distribution as shown in 
Fig.\ 9, but here each panel is limited to single stars
within a specified mass range. The top panel is for the high mass group
($8.5 M_\odot < M \leq 16 M_\odot$, the darkest shaded region in Fig.\ 7);
the middle panel $-$ the middle mass group ($4 M_\odot < M \leq 8.5 
M_\odot$); the bottom panel $-$ the low mass group ($2.5 M_\odot < 
M \leq 4 M_\odot$, the lightest shaded region in Fig.\ 7).}
\label{fig10}
\end{figure}

\placefigure{fig10} 

Surprisingly, the middle mass and low mass groups show much steeper
spin down slopes (with the interesting exception of the rapid rotator,
NGC~7160 \#940 = BD$+61^\circ 2222$, at $\log g_{\rm polar}=3.7$ 
in the bottom panel of Figure~10).    
Similar spin down differences among these mass groups were found for
the field B-stars by \citet{abt02}.  
This difference might imply that an additional 
angular momentum loss mechanism, perhaps involving magnetic fields, 
becomes important in these middle and lower mass B-type stars.
The presence of low gravity stars in the lower mass groups 
(summarized in Table~6) is puzzling if they are evolved objects. 
Our sample is drawn from young clusters (most are younger than 
$\log {\rm age} = 7.4$, except for NGC~2422 with $\log {\rm age} =7.86$;  
see Paper~I), so we would not expect to find any evolved 
stars among the objects in the middle and lower mass groups.   
\citet{mas95} present HR-diagrams for a number of young clusters, and they 
find many cases where there are significant numbers of late type 
B-stars with positions well above the main sequence.   
They argue that these objects are pre-main sequence stars
that have not yet contracted to their main sequence radii. 
We suspect that many of the low gravity objects shown in 
the middle and bottom panels of Figure~10
are also pre-main sequence stars.  If so, then they will evolve in the future  
from low to high gravity as they approach the main sequence, and our 
results would then suggest that they spin up as they do so to 
conserve angular momentum. 

\placetable{tab6}      

Our sample of 78 single-lined binary stars is too small to divide
into different mass groups, so we plot them all in one diagram in Figure~11. 
The binary systems appear to experience more spin down than the 
single B-stars (compare with Fig.~9).  
\citet{abt02} found that synchronization processes  
in short-period binary systems can dramatically reduce the rotational
velocity of the components.  If this is the major reason 
for the decline in $V \sin i$ in our binary sample, then 
it appears that tidal synchronization becomes significant in 
many close binaries when the more massive component attains 
a polar gravity of $\log g_{\rm polar}=3.9$, i.e., at a point when 
the star's larger radius makes the tidal interaction more 
effective in driving the rotation towards orbital synchronization. 

\begin{figure}
\plotone{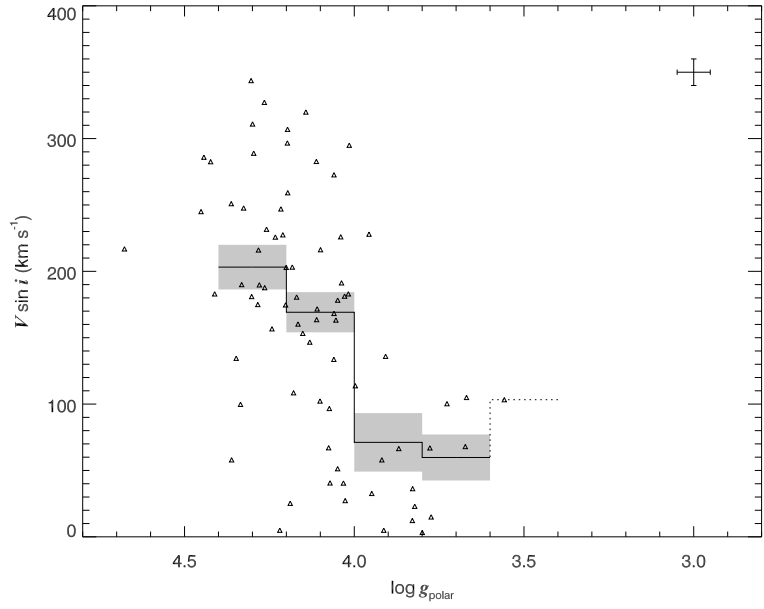}
\caption{
The $(\log g_{\rm polar}, V \sin i)$ distribution for probable 
single-lined binary systems.}
\label{fig11}
\end{figure}

\placefigure{fig11} 


\section{Helium Abundance}                          

Rotation influences the shape and temperature of a star's outer 
layers, but it also affects a star's interior structure.  
Rotation will promote internal mixing processes which
cause an exchange of gas between the core and the envelope, so that 
fresh hydrogen can migrate down to the core and fusion products can
be dredged up to the surface.  The consequence of this mixing is a gradual
abundance change of some elements on surface during the MS phase
(He and N become enriched while C and O decrease). 
The magnitude of the abundance change is predicted to be
related to stellar rotational velocity because faster rotation 
will trigger stronger mixing \citep{heg00,mey00}.  In this 
section we present He abundance measurements from our spectra 
that we analyze for the expected correlations with evolutionary
state and rotational velocity. 
  
\subsection{Measuring the Helium Abundance}

We can obtain a helium abundance by comparing the observed and model profiles
provided we have reliable estimates of $T_{\rm eff}$, $\log g$, and the
microturbulent velocity $V_t$.  We already have surface mean values for the 
first two parameters ($T_{\rm eff}$ and $\log g$) from H$\gamma$ line fitting (\S2). 
We adopted a constant value for the microturbulent velocity, $V_t = 2$ km~s$^{-1}$, 
that is comparable to the value found in multi-line studies of similar 
field B-stars \citep*{lyu04}.   The consequences of this simplification 
for our He abundance measurements are relatively minor.  
For example, we calculated the equivalent widths of 
\ion{He}{1} $\lambda\lambda 4026, 4387, 4471$ using a range of assumed 
$V_t = 0 - 8$ km~s$^{-1}$ for cases of $T_{\rm eff} = 16000$ and 20000~K and
$\log g = 3.5$ and 4.0.  The largest difference in the resulting 
equivalent width is $\approx 2.5\%$ between $V_t = 0$ and
8 km~s$^{-1}$ for the case of the \ion{He}{1} $\lambda 4387$ line 
at $T_{\rm eff} = 20000$~K and $\log g = 3.5$.  These \ion{He}{1} 
strength changes with microturbulent velocity are similar to the case 
presented by \citet{lyu04} for $T_{\rm eff} = 25000$~K and $\log g = 4.0$.
All of these results demonstrate that the changes in equivalent width 
of \ion{He}{1} $\lambda\lambda 4026, 4387, 4471$ that result from a different
choice of $V_t$ are small compared to observational errors for MS 
B-type stars.  The $V_t$ measurements of field B stars by \citet{lyu04} are mainly
lower than 8 km~s$^{-1}$ with a few exceptions of hot and evolved stars,
which are rare in our sample.  Thus, our assumption of constant  
$V_t = 2$ km~s$^{-1}$ for all the sample stars will have a negligible 
impact on our derived helium abundances.

The theoretical \ion{He}{1} $\lambda\lambda 4026, 4387, 4471$ profiles were calculated 
using the SYNSPEC code and Kurucz line blanketed LTE atmosphere models in same 
way as we did for the H$\gamma$ line (\S2) to include rotational and 
instrumental broadening.  We derived five template spectra 
for each line corresponding to helium abundances of 1/4, 1/2, 1, 2, and 
4 times the solar value.  We then made a bilinear interpolation in our 
$(T_{\rm eff}, \log g)$ grid to estimate the profiles over the run of 
He abundance for the specific temperature and gravity of each star.  
The $\chi^2$ residuals of the differences between each of the five template 
and observed spectra were fitted with a polynomial
curve to locate the minimum residual position and hence the He 
abundance measurement for the particular line.  
Examples of the fits are illustrated in Figure~12.  Generally each 
star has three abundance measurements from the three \ion{He}{1} lines, 
and these are listed in Table~1 (columns 9 -- 11) together with the mean 
and standard deviation of the He abundance (columns 12 -- 13).
(All of these abundances include a small correction for non-LTE 
effects that is described in the next paragraph.)   
Note that one or more measurements may be missing for some stars due to: 
(1) line blending in double-lined spectroscopic binaries;
(2) excess noise in the spectral regions of interest; 
(3) severe line blending with nearby metallic transitions;
(4) extreme weakness of the \ion{He}{1} lines in the cooler stars ($T_{\rm eff} < 11500$~K);
(5) those few cases where the He abundance appears to be either extremely high
($\gg 4\times$ solar) or low ($\ll 1/4\times$ solar) and beyond the scope of our abundance analysis.
We show examples of a He-weak and a He-strong spectrum in Figure~13.  
These extreme targets are the He-weak stars IC~2395 \#98, \#122, NGC~2244 \#59, \#298, and
NGC~2362 \#73, and the He-strong star NGC~6193 \#17. 

\begin{figure}
\plotone{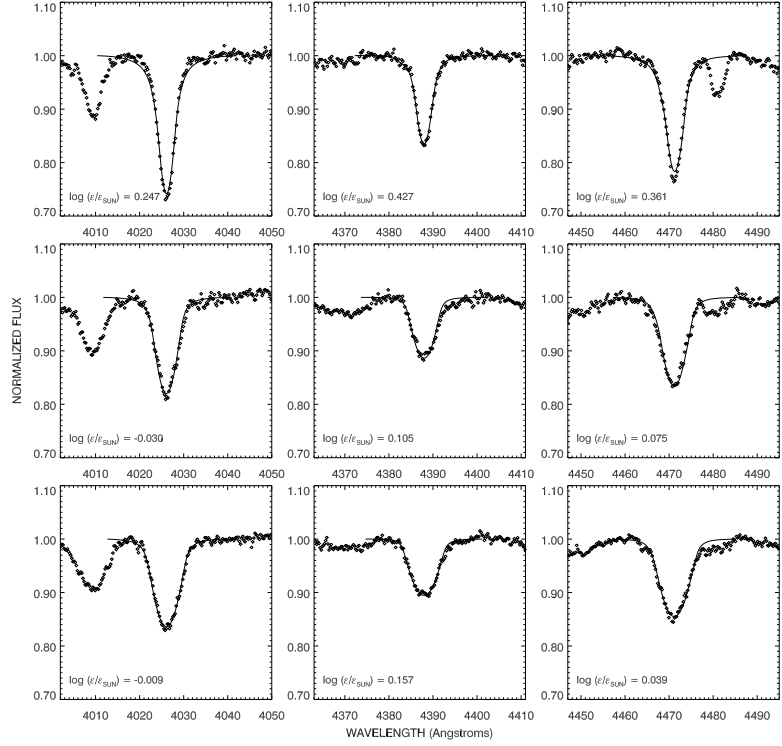}
\caption{Line profile fitting results made to determine the He abundance
for the same three stars used in Fig.\ 3 and 4.  Each panel lists 
the derived LTE He abundance relative to the solar value for 
\ion{He}{1} $\lambda 4026$ ({\it left column}), 
\ion{He}{1} $\lambda 4387$ ({\it middle column}), and 
\ion{He}{1} $\lambda 4471$ ({\it right column}).  
The stars are ({\it from top to bottom rows})  
NGC~1502 \#23, NGC~869 \#803, and NGC~884 \#2255.}
\label{fig12}
\end{figure}

\placefigure{fig12} 

\begin{figure}
\plotone{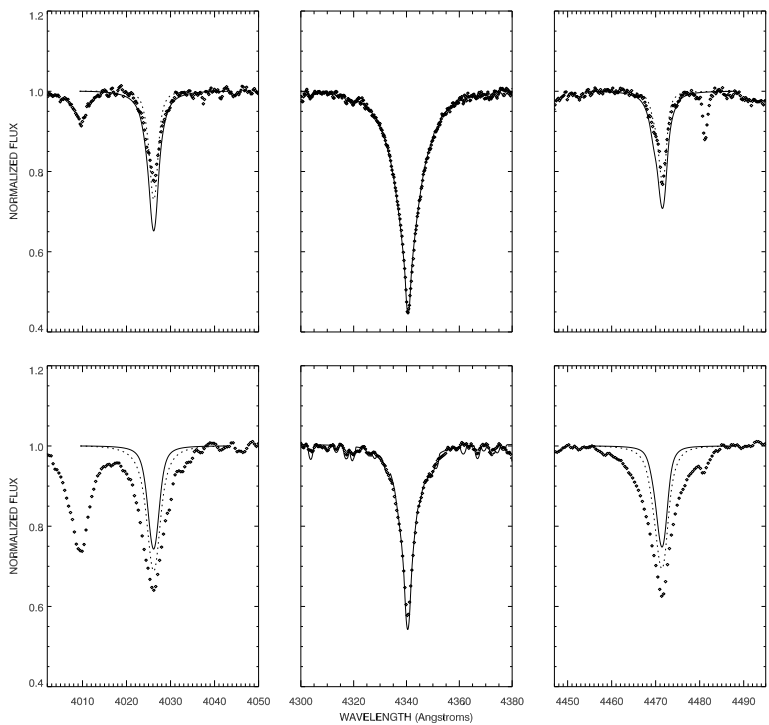}   
\caption{Top panel: An example of a He-weak star, NGC~1502 \#44. The 
panels from left to right: the \ion{He}{1} $\lambda 4026$ region, 
the H$\gamma$ region, and the \ion{He}{1} $\lambda 4471$ region. 
The solid lines are LTE theoretical profiles synthesized assuming 
solar abundances and based upon the derived $T_{\rm eff}$, 
$\log g$, and $V \sin i$ for the star while the dotted lines
are LTE theoretical profiles synthesized assuming 1/4 the solar He abundance.
Bottom panel: An example of an extreme He-strong star, NGC~6193 \#17
(= CD$-48^\circ11051$; first noted as a He-rich star by \citealt{arn88}).
The dotted lines are theoretical profiles synthesized assuming 4 times 
the solar He abundance.}
\label{fig13}
\end{figure}

\placefigure{fig13} 

The He abundances we derive are based on LTE models for H and He which may not
be accurate due to neglect of non-LTE effects, especially for hot and more evolved B 
giant stars ($T_{\rm eff} > 25000$~K and $\log g < 3.5 $).  We need to apply some reliable 
corrections to the abundances to account for these non-LTE effects. The differences in the
He line equivalent widths between LTE and non-LTE models were investigated by \citet{aue73}.
However, their work was based on simple atmosphere models without line blanketing, 
and thus, is not directly applicable for our purposes.  Fortunately, we were
able to obtain a set of non-LTE, line-blanketed model B-star spectra from
T.\ Lanz and I.\ Hubeny \citep{lan05} that represent an extension of their OSTAR2002 
grid \citep{lan03}.  We calculated the equivalent widths for both the LTE (based on 
Kurucz models) and non-LTE (Lanz \& Hubeny) cases (see Table~7), and then used
them to derive He abundance ($\epsilon$) corrections based upon stellar temperature 
and gravity.  These corrections are small in most cases ($\Delta\log (\epsilon) < 0.1$ dex) 
and are only significant among the hotter and more evolved stars.  

\placetable{tab7}      

The He abundances derived from each of the three \ion{He}{1} lines 
should lead to a consistent result in principle.  
However, \citet{lyu04} found that line-to-line differences do exist. 
They showed that the ratio of the equivalent width of \ion{He}{1} $\lambda 4026$
to that of \ion{He}{1} $\lambda 4471$ decreases with increasing 
temperature among observed B-stars, while theoretical models 
predict a constant or increasing ratio between these lines 
among the hotter stars (and a similar trend exists between the 
\ion{He}{1} $\lambda 4387$ and \ion{He}{1} $\lambda 4922$ 
equivalent widths).  The direct consequence of this discrepancy is that the He
abundances derived from \ion{He}{1} $\lambda\lambda 4471,4922$ are greater than 
those derived from \ion{He}{1} $\lambda\lambda 4026,4387$.
The same kind of line-to-line differences are apparently present 
in our analysis as well.   We plot in Figure~14 the derived 
He abundance ratios $\log [\epsilon(4471)/\epsilon(4026)]$ 
and $\log [\epsilon(4387)/\epsilon(4026)]$ as a function of $T_{\rm eff}$.
The mean value of $\log [\epsilon(4471)/\epsilon(4026)]$ increases
from $\approx 0.0$ dex at the cool end to +0.2 dex at $T_{\rm eff} = 26000$~K. 
On the other hand, the differences between the abundance results
from \ion{He}{1} $\lambda 4026$ and \ion{He}{1} $\lambda 4387$ 
are small except at the cool end where they differ roughly by +0.1 dex 
(probably caused by line blending effects from the neglected lines of 
\ion{Mg}{2} $\lambda 4384.6, 4390.6$ and \ion{Fe}{2} $\lambda 4385.4$
that strengthen in the cooler B-stars).  \citet{lyu04} advocate the 
use of the \ion{He}{1} $\lambda\lambda 4471,4922$ lines based upon 
their better broadening theory and their consistent results 
for the helium weak stars.  Because our data show similar
line-to-line differences, we will focus our attention on the 
abundance derived from \ion{He}{1} $\lambda 4471$, as advocated by \citet{lyu04}.  
Since both the individual and mean line abundances are given in 
Table~1, it is straight forward to analyze the data for any subset of these lines.    

\begin{figure}
\plotone{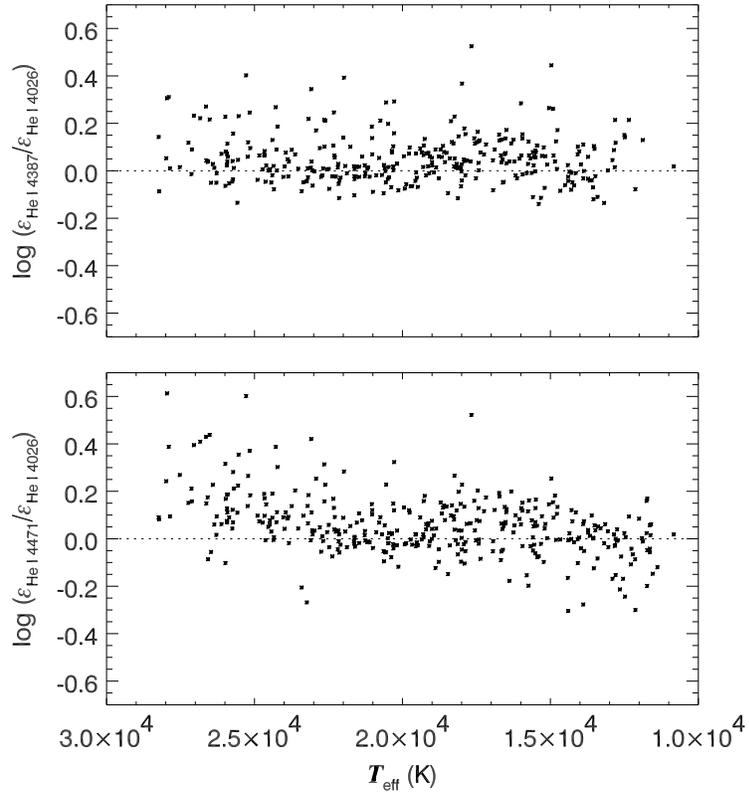}                   
\caption{The difference in the non-LTE corrected He abundance from three  
different He lines plotted against $T_{\rm eff}$.  The top panel shows 
the differences between the results from 
\ion{He}{1} $\lambda 4387$ and \ion{He}{1} $\lambda 4026$
and the bottom panel shows the differences between 
\ion{He}{1} $\lambda 4471$ and \ion{He}{1} $\lambda 4026$.
}
\label{fig14}
\end{figure}

\placefigure{fig14} 

We used the standard deviation of the He abundance measurements from 
\ion{He}{1} $\lambda\lambda 4026$, $4387$, $4471$ as a measure of the 
He abundance error (last column of Table~1), which is adequate  
for statistical purposes but may underestimate the actual errors in some cases. 
The mean errors in He abundance are 
$\pm0.07$~dex for stars with $T_{\rm eff} \geq 23000$~K, 
$\pm0.04$~dex for stars with $23000 {\rm ~K} > T_{\rm eff} \geq 17000$~K, and 
$\pm0.05$~dex for stars with $T_{\rm eff} < 17000$~K.   
These error estimates reflect both the noise in the observed spectra and the line-to-line
He abundance differences discussed above.  The errors in the He abundance due to 
uncertainties in the derived $T_{\rm eff}$ and $\log g$ values (columns 4 and 6  
of Table~1) and due to possible differences in microturbulence from the adopted 
value are all smaller ($<0.04$ dex) than means listed above. 

\subsection{Evolution of the Helium Abundance}

We plot in Figure~15 our derived He abundances for all the 
single stars and single-lined binaries in our sample versus $\log g_{\rm polar}$, 
which we suggest is a good indicator of evolutionary state (\S3). 
The scatter in this figure is both significant (see the error 
bar in the upper left hand corner) and surprising. 
There is a concentration of data points near the solar He abundance 
that shows a possible trend of increasing He abundance with age
(and decreasing $\log g_{\rm polar}$), but a large fraction 
of the measurements are distributed over a wide range in He 
abundance.  Our sample appears to contain a large number of 
helium peculiar stars, both weak and strong, in striking 
contrast to the sample analyzed by \citet{lyu04} who identified 
only two helium weak stars out of 102 B0 - B5 field stars.  
Any evolutionary trend of He abundance that may exist in Figure~15
is lost in the large scatter introduced by the He peculiar stars. 

\begin{figure}
\plotone{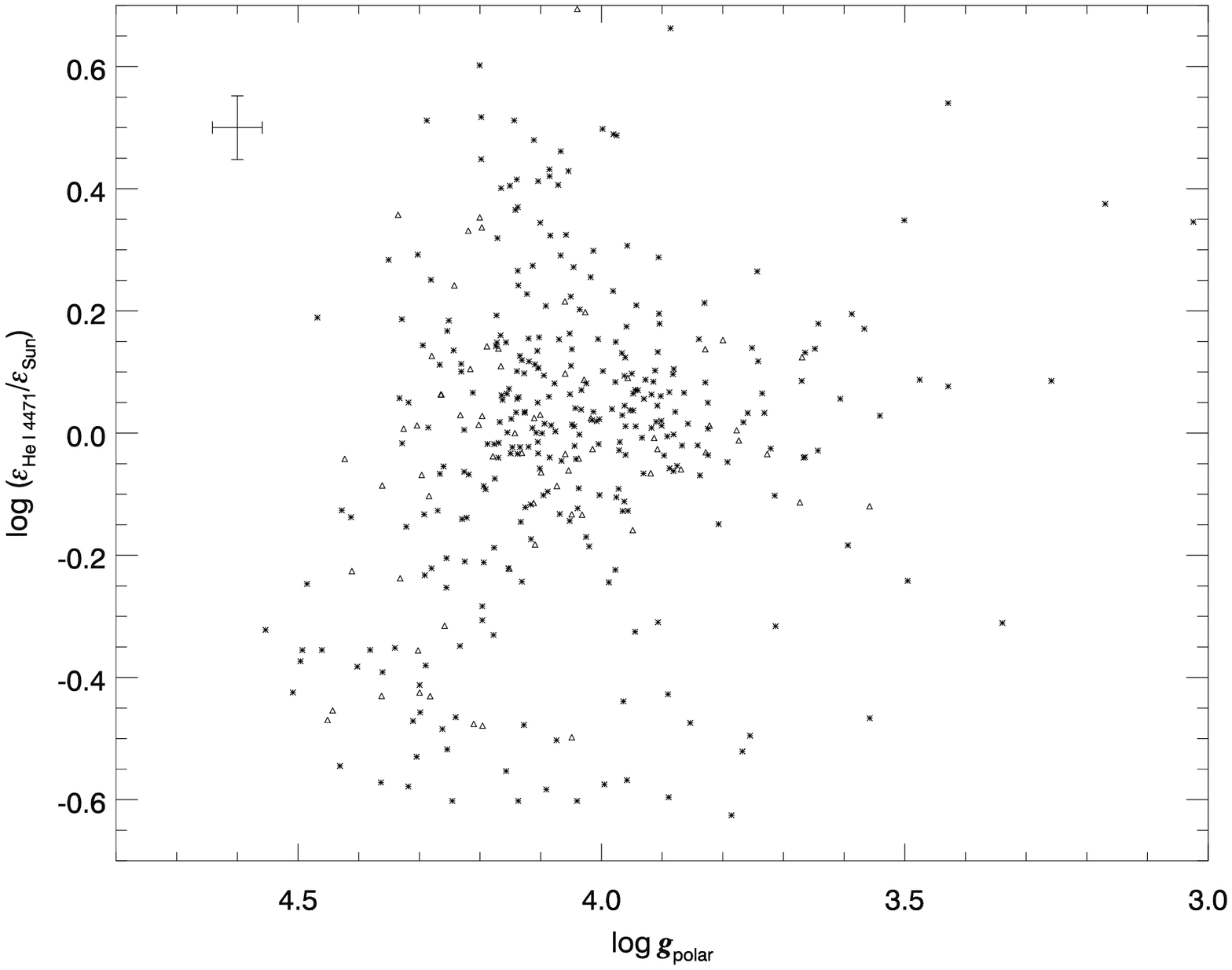}
\caption{A plot of the non-LTE corrected He abundance versus $\log g_{\rm polar}$
for the whole sample.  The average measurement errors are indicated 
in the upper-left corner. Asterisks are assigned to single stars 
($N = 325$) while
triangles are used for binary systems ($N = 78$).} 
\label{fig15}
\end{figure}

\placefigure{fig15} 

Studies of the helium peculiar stars \citep*{bor79,bor83} 
indicate that they are found only among stars of subtype  
later than B2.  This distribution is clearly confirmed 
in our results.  We plot in Figure~16 a diagram of 
He abundance versus $T_{\rm eff}$, where we see that almost
all the He peculiar stars have temperatures $T_{\rm eff} < 23000$~K.
Below 20000~K (B2) we find that about one third (67 of 199) 
of the stars have a large He abundance deviation,  
$|\log (\epsilon/\epsilon_\odot)| > 0.3$ dex, while only 
8 of 127 stars above this temperature have such He peculiarities. 
In fact, in the low temperature range ($T_{\rm eff} < 18000$~K), 
the helium peculiar stars are so pervasive and uniformly 
distributed in abundance that there are no longer any 
clear boundaries defining the He-weak, He-normal, and He-strong stars 
in our sample.   The mean observational errors in abundance are much 
smaller than the observed spread in He abundance seen in Figure~16. 

\begin{figure}
\plotone{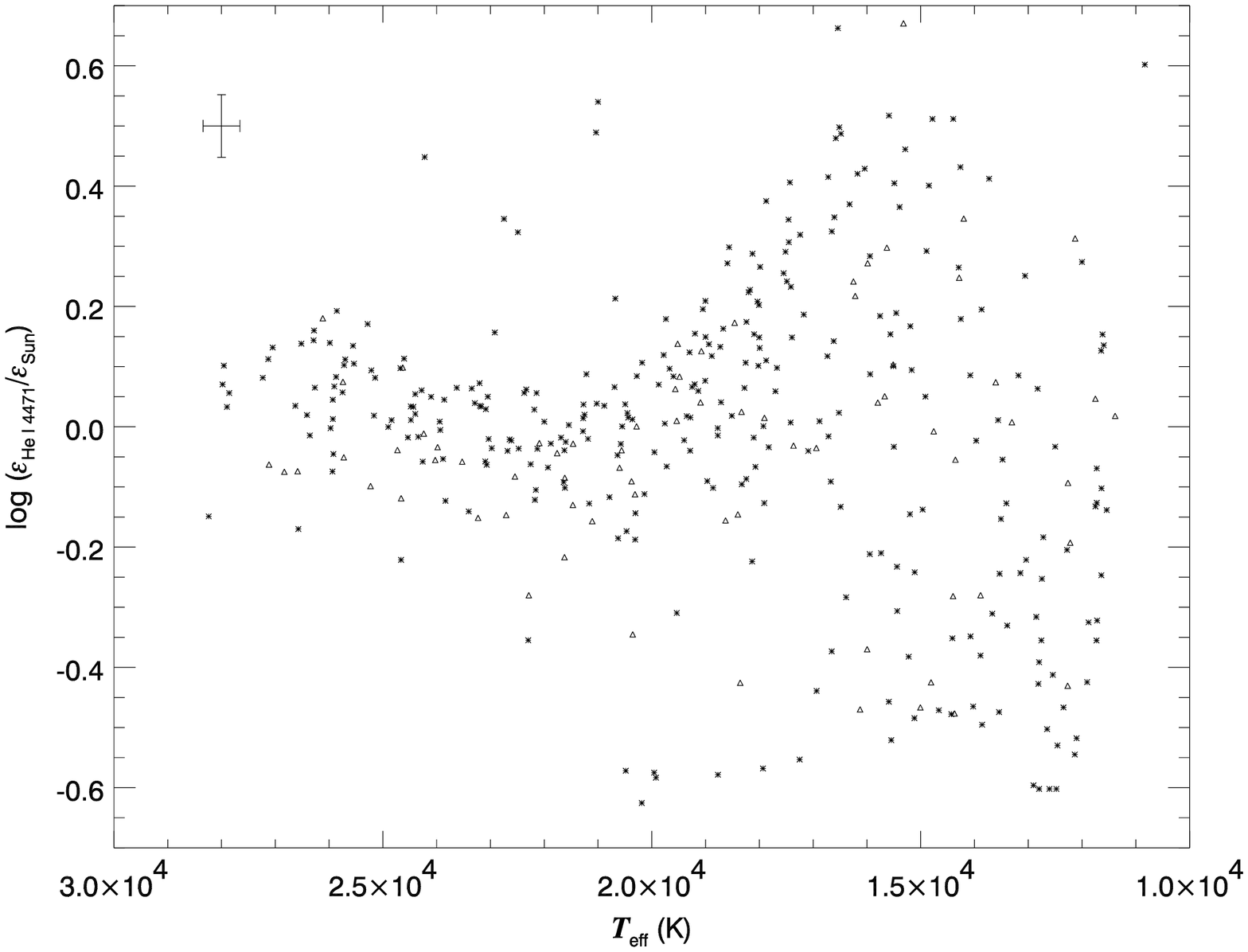}   
\caption{A plot of the non-LTE corrected He abundance versus $T_{\rm eff}$ 
for the whole sample. The average measurement errors are indicated 
in the upper-left corner. Asterisks are assigned to single stars 
($N = 325$) while
triangles are used for binary systems ($N = 78$).}
\label{fig16}
\end{figure}

\placefigure{fig16} 

There is much evidence to suggest that both the He-strong and He-weak 
stars have strong magnetic fields that alter the surface abundance  
distribution of some chemical species \citep{mat04}. 
Indeed there are some helium variable stars, such as HD~125823, that
periodically vary between He-weak and He-strong over 
their rotation cycle \citep{jas68}.  Because of the preponderance
of helium peculiar stars among the cooler objects in our sample, 
we cannot easily differentiate between helium enrichment 
due to evolutionary effects or due to magnetic effects. 
Therefore, we will restrict our analysis of evolutionary effects
to those stars with $T_{\rm eff} > 23000$~K where no 
He peculiar stars are found.   This temperature range 
corresponds approximately to the high mass group of single stars 
(88 objects) plotted in the darker shaded region of Figure~7.

The new diagram of He abundance versus $\log g_{\rm polar}$ for the high mass 
star group ($8.5 M_\odot < M < 16 M_\odot$) appears in Figure~17.  
We can clearly see in this figure that the surface helium abundance is 
gradually enriched as stars evolve from ZAMS ($\log g_{\rm polar} = 4.3$) to TAMS
($\log g_{\rm polar} = 3.5$).  We made a linear least squares
fit to the data (shown as a dotted line)
\begin{displaymath}
\log (\epsilon/\epsilon_\odot) = (-0.114\pm0.059)~\log g_{\rm polar}~+~(0.494\pm0.012)
\end{displaymath}
that indicates an average He abundance increase of   
$0.09\pm0.05$ dex (or $23\pm13 \%$) between 
ZAMS ($\log g_{\rm polar} = 4.3$) and TAMS ($\log g_{\rm polar} = 3.5$). 
This estimate is in reasonable agreement with the results of 
\citet{lyu04} who found a ZAMS to TAMS increase 
in He abundance of $26\%$ for stars in the mass range $4 - 11 M_\odot$
and $67\%$ for more massive stars in the range $12 - 19 M_\odot$. 

\begin{figure}
\plotone{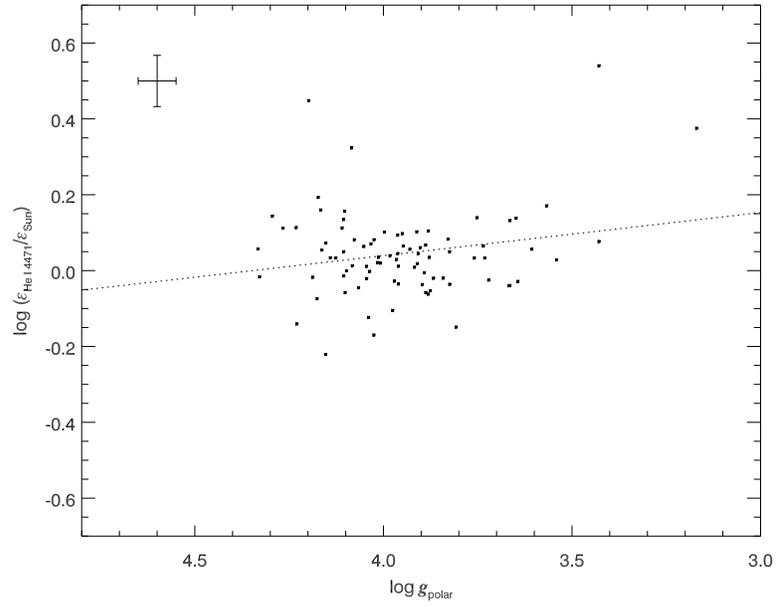}   
\caption{A plot of the non-LTE corrected He abundance versus $\log g_{\rm polar}$
for the high mass group of single stars (see the darker shaded area in
Fig.~7).  The average measurement errors are indicated in
the upper-left corner.  The dotted line is the linear least squares
fit.}
\label{fig17}
\end{figure}

\placefigure{fig17} 

\subsection{Rotational Effects on the Helium Abundance}

The theoretical models for mixing in rotating stars 
predict that the enrichment of surface helium increases with 
age and with rotation velocity.  The faster the stars rotate, 
the greater will be the He enrichment as stars evolve towards the TAMS. 
In order to search for a correlation between He abundance 
and rotation ($V\sin i$), we must again restrict our sample to the 
hotter, more massive stars to avoid introducing the complexities of 
the helium peculiar stars (\S5.2). 

If the He abundance really does depend on both evolutionary 
state and rotational velocity, then it is important to 
select subsamples of comparable evolutionary status in 
order to investigate how the He abundances may vary 
with rotation.   We divided the same high mass group (\S5.2) into three 
subsamples according to their $\log g_{\rm polar}$ values, namely 
the young subgroup (22 stars, $4.5 \geq \log g_{\rm polar} > 4.1$), 
the mid-age subgroup (47 stars, $4.1 \geq \log g_{\rm polar} > 3.8$), and 
the old subgroup (14 stars, $3.8 \geq \log g_{\rm polar} > 3.4$). 
We plot the distribution of He abundance versus 
$V \sin i$ for these three subgroups in the three panels of Figure~18. 
Because each panel contains only stars having
similar evolutionary status (with a narrow range in $\log g_{\rm polar}$), 
the differences in He abundance due to differences in evolutionary state
are much reduced.  Therefore, any correlation between He abundance
and $V \sin i$ found in each panel will reflect mainly the influence of
stellar rotation.   We made linear least squares fits for each of these
subgroups that are also plotted in each panel.  
The fit results are (from young to old):
\begin{displaymath}
\log (\epsilon/\epsilon_\odot) = (-0.0\pm2.5)\times10^{-4}~V \sin i~+~(0.043\pm0.024)
\end{displaymath}
\begin{displaymath}
\log (\epsilon/\epsilon_\odot) = (0.3\pm1.6)\times10^{-4}~V \sin i~+~(0.015\pm0.013)
\end{displaymath}
\begin{displaymath}
\log (\epsilon/\epsilon_\odot) = (4.1\pm2.3)\times10^{-4}~V \sin i~+~(0.009\pm0.019)
\end{displaymath}
We can see that there is basically no dependence on rotation for the He
abundances of the stars in the young and mid-age subgroups.  
However, there does appear to be a correlation between He abundance 
and rotation among the stars in the old subgroup.  Though there are fewer 
stars with high $V \sin i$ in the old group (perhaps due to spin down), 
a positive slope is clearly seen that is larger than that of 
the younger subgroups.  

\begin{figure}
\plotone{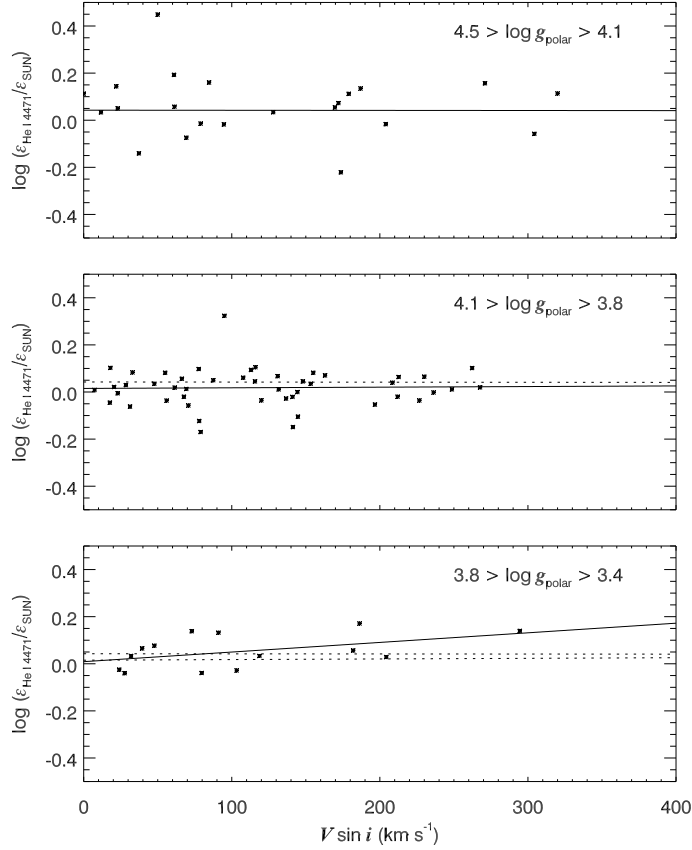}
\caption{The stars shown in Fig.~17 are divided into 
three subgroups by their $\log g_{\rm polar}$ values, and plotted
in three panels of the non-LTE corrected He abundance versus $V \sin i$. The youngest subgroup
is plotted in the top panel while the oldest subgroup is plotted in the bottom panel. 
The solid lines are the linear least squares fit for each subgroup. 
The dotted lines are the fits for the younger subgroups replotted 
in each older group for comparison purposes.}
\label{fig18}
\end{figure}

\placefigure{fig18} 

Our results appear to support the predictions for the evolution
of rotating stars, specifically that rotationally induced 
mixing during the MS results in a He enrichment of the 
photosphere (Fig.~17) and that the enrichment is greater in 
stars that spin faster (Fig.~18).   The qualitative agreement 
is gratifying, but it is difficult to make a quantitative 
comparison with theoretical predictions because our 
rotation measurements contain the unknown projection factor $\sin i$ 
and because our samples are relatively small.  However, 
both problems will become less significant as more observations 
of this kind are made. 


\section{Conclusions}                             

Our main conclusions can be summarized as follows:

(1) We determined average effective temperatures ($T_{\rm eff}$) 
and gravities ($\log g$) of 461 OB stars in 19 young clusters 
(most of which are MS stars) by fitting the H$\gamma$ profile
in their spectra.  Our numerical tests using realistic 
models for rotating stars show that the measured $T_{\rm eff}$ and
$\log g$ are reliable estimates of the average physical conditions 
in the photosphere for most of the B-type stars we observed. 

(2) We used the profile synthesis results for rotating stars to
develop a method to estimate the true polar gravity of a rotating star
based upon its measured $T_{\rm eff}$, $\log g$, and $V \sin i$. 
We argue that $\log g_{\rm polar}$ is a better indicator of the evolutionary
status of a rotating star than the average $\log g$ (particularly 
in the case of rapid rotators). 

(3) A statistical analysis of the $V\sin i$ distribution as 
a function of evolutionary state ($\log g_{\rm polar}$) shows that all these OB
stars experience a spin down during the MS phase as theories of rotating stars
predict.  The spin down behavior of the high mass star group 
in our sample ($8.5 M_\odot < M < 16 M_\odot$) quantitatively
agrees with theoretical calculations that assume that the spin 
down is caused by rotationally-aided stellar wind mass loss. 
We found a few relatively fast rotators among stars nearing the TAMS, 
and these may be stars spun up by a short core contraction phase or by mass 
transfer in a close binary.  We also found that close binaries 
generally experience a significant spin down around the stage where 
$\log g_{\rm polar} = 3.9$ that is probably the result of 
tidal interaction and orbital synchronization.  

(4) We determined He abundances for most of the stars through 
a comparison of the observed and synthetic profiles of \ion{He}{1} lines. 
Our non-LTE corrected data show that the He abundances 
measured from \ion{He}{1} $\lambda 4026$ and 
from \ion{He}{1} $\lambda 4471$ differ by a small amount that
increases with the temperature of the star (also found 
by \citealt{lyu04}). 

(5) We were surprised to find that our sample contains many
helium peculiar stars (He-weak and He-strong), which are mainly
objects with $T_{\rm eff} < 23000$~K.  In fact, the distribution 
of He abundances among stars with $T_{\rm eff} < 18000$~K
is so broad and uniform that it becomes difficult to 
differentiate between the He-weak, He-normal, and
He-strong stars.  Unfortunately, this scatter makes impossible 
an analysis of evolutionary He abundance changes for the 
cooler stars. 

(6) Because of the problems introduced by the large number 
of helium peculiar stars among the cooler stars, 
we limited our analysis of evolutionary changes in the 
He abundance to the high mass stars.  We found that the
He abundance does increase among stars of more advanced 
evolutionary state (lower $\log g_{\rm polar}$) and, 
within groups of common evolutionary state, among stars 
with larger $V\sin i$.  This analysis supports the theoretical 
claim that rotationally induced mixing plays a key role in the 
surface He enrichment of rotating stars. 

(7) The lower mass stars in our sample have two remarkable 
properties: relatively low spin rates among the lower gravity stars  
and a large population of helium peculiar stars.  We suggest that 
both properties may be related to their youth.  
The lower gravity stars are probably pre-main sequence objects 
rather than older evolved stars, and they are destined to  
spin up as they contract and become main sequence stars. 
Many studies of the helium peculiar stars \citep{bor79,bor83,wad97,sho04} 
have concluded that they have strong magnetic fields 
which cause a non-uniform distribution of helium in 
the photosphere.  We expect that many young B-stars are 
born with a magnetic field derived from their natal cloud, so 
the preponderance of helium peculiar stars among the
young stars of our sample probably reflects the relatively 
strong magnetic fields associated with the newborn stars. 
  

\acknowledgments

We are grateful to the KPNO and CTIO staffs and especially
Diane Harmer and Roger Smith for their help in making these
observations possible.  We would like to thank Richard Townsend
and Paul Wiita for their very helpful comments.  We also 
especially grateful to Ivan Hubeny and Thierry Lanz for their 
assistance with the TLUSTY and SYNSPEC codes and for 
sending us their results on the non-LTE atmospheres and 
spectra of B-stars in advance of publication.  
This material is based on work supported by the National Science
Foundation under Grant No.~AST-0205297.
Institutional support has been provided from the GSU College
of Arts and Sciences and from the Research Program Enhancement
fund of the Board of Regents of the University System of Georgia,
administered through the GSU Office of the Vice President for Research.
We gratefully acknowledge all this support.





\clearpage

\clearpage
         
\clearpage
         
\clearpage
         
\clearpage

\clearpage

\clearpage

\clearpage

\clearpage

\clearpage

\clearpage

\clearpage

\clearpage   

\clearpage

\clearpage

\clearpage

\clearpage 

\clearpage

\clearpage   


\begin{deluxetable}{lcccccccccccc}
\rotate
\tabletypesize{\footnotesize}
\tablewidth{0pc}
\tablecaption{Stellar Physical Parameters\tablenotemark{a}
\label{tab1}}
\tablehead{
\colhead{Cluster } &
\colhead{WEBDA  } &
\colhead{$T_{\rm eff}$ } &
\colhead{$\Delta T_{\rm eff}$   } &
\colhead{ } &
\colhead{ } &
\colhead{$V \sin i$ } &
\colhead{} &
\colhead{$\log[{\epsilon\over\epsilon_\odot}]$} &
\colhead{$\log[{\epsilon\over\epsilon_\odot}]$} &
\colhead{$\log[{\epsilon\over\epsilon_\odot}]$} &
\colhead{ } &
\colhead{ } \\
\colhead{Name } &
\colhead{Index } &
\colhead{(K) } &
\colhead{(K)  } &
\colhead{$\log g$ } &
\colhead{$\Delta\log g$ } &
\colhead{(km~s$^{-1})$ } &
\colhead{$\log g_{\rm polar}$} &
\colhead{(4026\AA )} &
\colhead{(4387\AA )} &
\colhead{(4471\AA )} &
\colhead{$\overline{\log[{\epsilon\over\epsilon_\odot}]}$} &
\colhead{$\Delta\log[{\epsilon\over\epsilon_\odot}]$}
}
\startdata
Ber86\dotfill&\phn1& 26878 & 406 & 3.84 & 0.04 &  185 & 3.93 & $-0.02$ & $-0.05$ &\phs0.22 &\phs0.05 & 0.12 \\
Ber86\dotfill&  12 & 15440 & 232 & 4.05 & 0.04 &  309 & 4.29 & $-0.30$ & $-0.09$ & $-0.39$ & $-0.26$ & 0.12 \\
Ber86\dotfill&  13 & 28711 & 191 & 3.59 & 0.02 &  183 & 3.73 & $-0.11$ & $-0.19$ &\phs0.26 & $-0.02$ & 0.20 \\
Ber86\dotfill&  15 & 23049 & 316 & 4.10 & 0.03 &\phn23& 4.10 &\phs0.04 &\phs0.01 &\phs0.10 &\phs0.05 & 0.04 \\
\enddata
\tablenotetext{a}{The full version of this table appears in the on-line edition.}
\end{deluxetable}

\clearpage


\begin{deluxetable}{lcl}
\tabletypesize{\scriptsize}
\rotate
\tablewidth{0pc}
\tablecaption{Photometric Data Sources for Cluster Stars
\label{tab2}}
\tablehead{
\colhead{Cluster  } &
\colhead{$E(B-V)$  } &
\colhead{  } \\
\colhead{Name  } &
\colhead{\citep{lok01} } &
\colhead{Sources }
}

\startdata
NGC 6193   \dotfill &    0.475 &    \citet{her77,vaz91} \\
Trumpler 16\dotfill &    0.493 &    \citet*{fei69,fei73,fei82}; \\
                    &          &    \citet{mas93,tap03} \\
IC 1805    \dotfill &    0.822 &    \citet*{jos83,mas95} \\
IC 2944    \dotfill &    0.320 &    \citet{tha65,ard77,ped84} \\
Trumpler 14\dotfill &    0.516 &    \citet{fei73,tap03} \\
NGC 2244   \dotfill &    0.463 &    \citet{joh62,tur76,ogu81};\\
                    &          &    \citet{per87,mas95} \\
NGC 2384   \dotfill &    0.255 &    \citet{vog72} \\
NGC 2362   \dotfill &    0.095 &    \citet{joh53,per73} \\
NGC 3293   \dotfill &    0.263 &    \citet{tur80,her82,bau03} \\
NGC 884    \dotfill &    0.560 &    \citet*{tap84,sle02} \\
NGC 1502   \dotfill &    0.759 &    \citet{dom64,gue64}; \\
                    &          &    \citet{pur64,rei87} \\
NGC 869    \dotfill &    0.575 &    \citet{joh55,hil56,sch65}; \\
                    &          &    \citet{tap84,sle02} \\
NGC 2467   \dotfill &    0.338 &    \citet{lod66,fei89} \\
Berkeley 86\dotfill &    0.898 &    \citet{for92,mas95} \\
IC 2395    \dotfill &    0.066 &    \citet{clar03} \\
NGC 4755   \dotfill &    0.388 &    \citet{per76,dac84,san01} \\
NGC 7160   \dotfill &    0.375 &    \citet{hil77,car82,deg83} \\
NGC 457    \dotfill &    0.472 &    \citet{pes59,hoa61,mof74} \\
NGC 2422   \dotfill &    0.070 &    \citet*{lyn59,deu76,pri03} \\
\enddata
\end{deluxetable}

\clearpage


\begin{deluxetable}{lccccc}
\tablewidth{0pc}
\tablecaption{Physical Parameters of Models
\label{tab3}}
\tablehead{
\colhead{Model    } &
\colhead{$M$ } &
\colhead{$R_{\rm polar}$ } &
\colhead{$T_{\rm polar}$ } &
\colhead{ } &
\colhead{$V_c$ } \\
\colhead{Number   } &
\colhead{$(M_\odot)$ } &
\colhead{$(R_\odot)$ } &
\colhead{(K)   } &
\colhead{$\log g_{\rm polar}$  } &
\colhead{(km~s$^{-1})$ }
}
\startdata
1\dotfill & 9.5 &    4.0 & 25500 & 4.211 & 549 \\ 
2\dotfill & 9.5 &    5.0 & 25500 & 4.018 & 491 \\ 
3\dotfill & 9.5 &    6.4 & 25500 & 3.803 & 434 \\ 
4\dotfill & 5.5 &    2.9 & 18700 & 4.253 & 491 \\
5\dotfill & 5.5 &    3.9 & 18700 & 3.996 & 423 \\
6\dotfill & 5.5 &    4.9 & 18700 & 3.798 & 378 \\
7\dotfill & 4.0 &    2.7 & 15400 & 4.177 & 434 \\
8\dotfill & 4.0 &    3.4 & 15400 & 3.977 & 387 \\
9\dotfill & 4.0 &    4.1 & 15400 & 3.814 & 352 \\
\enddata
\end{deluxetable}
\clearpage


\begin{deluxetable}{lccccccccc}
\tabletypesize{\scriptsize}
\tablewidth{0pc}
\tablecaption{Test Results for Model \#1
\label{tab4}}
\tablehead{
\colhead{$i$ } &
\colhead{$V \sin i $ } &
\colhead{$T_{msr}$   } &
\colhead{  } &
\colhead{$T_{geo}$ } &
\colhead{ } & 
\colhead{$T_{flux}$} &
\colhead{ } & 
\colhead{$T_L$ } & 
\colhead{ }\\
\colhead{(deg) } &
\colhead{(km~s$^{-1}$) } &
\colhead{(K)   } &
\colhead{$\log g_{msr}$  } &
\colhead{(K) } & 
\colhead{$\log g_{geo}$} & 
\colhead{(K)} & 
\colhead{$\log g_{flux}$} &
\colhead{(K)} & 
\colhead{$\delta\log g$}}
\startdata
90  &\phn50  &  25437 & 4.208 & 25447 & 4.208 & 25446 & 4.208  & 25453 & 0.004 \\
90  &   100  &  25282 & 4.195 & 25288 & 4.197 & 25284 & 4.197  & 25312 & 0.016 \\
90  &   200  &  24656 & 4.143 & 24635 & 4.151 & 24625 & 4.151  & 24744 & 0.068 \\
90  &   300  &  23555 & 4.046 & 23489 & 4.068 & 23487 & 4.068  & 23790 & 0.165 \\
90  &   400  &  21927 & 3.885 & 21747 & 3.931 & 21834 & 3.938  & 22462 & 0.327 \\
70  &\phn50  &  25432 & 4.207 & 25443 & 4.207 & 25442 & 4.207  & 25447 & 0.004 \\
70  &   100  &  25262 & 4.195 & 25269 & 4.196 & 25267 & 4.195  & 25287 & 0.017 \\
70  &   200  &  24562 & 4.139 & 24559 & 4.146 & 24558 & 4.146  & 24644 & 0.072 \\
70  &   300  &  23392 & 4.044 & 23314 & 4.054 & 23350 & 4.057  & 23562 & 0.167 \\
70  &   400  &  21489 & 3.867 & 21423 & 3.902 & 21661 & 3.921  & 22072 & 0.344 \\
50  &\phn50  &  25407 & 4.206 & 25422 & 4.206 & 25423 & 4.206  & 25420 & 0.005 \\
50  &   100  &  25173 & 4.191 & 25188 & 4.190 & 25191 & 4.190  & 25179 & 0.020 \\
50  &   200  &  24232 & 4.126 & 24229 & 4.122 & 24267 & 4.125  & 24207 & 0.085 \\
50  &   300  &  22549 & 4.007 & 22540 & 3.992 & 22773 & 4.010  & 22586 & 0.204 \\
30  &\phn50  &  25323 & 4.203 & 25341 & 4.200 & 25346 & 4.201  & 25312 & 0.008 \\
30  &   100  &  24823 & 4.180 & 24859 & 4.167 & 24888 & 4.169  & 24744 & 0.032 \\
30  &   200  &  22631 & 4.052 & 22820 & 4.014 & 23117 & 4.037  & 22462 & 0.159 \\
\enddata
\end{deluxetable}

\clearpage


\begin{deluxetable}{ccccc}
\tablewidth{0pc}
\tablecaption{Polar Gravity Corrections
\label{tab5}}
\tablehead{
\colhead{Model    } &
\colhead{$V \sin i$   } &
\colhead{$T_{msr}$   } &
\colhead{  } &
\colhead{  } \\
\colhead{Number    } &
\colhead{(km~s$^{-1}$)   } &
\colhead{(K)   } &
\colhead{$\log g_{msr}$  } &
\colhead{$\delta \log g$ }}
\startdata
1 & \phn 50    &  25303   &   4.203  &  0.009  \\
1 & 100   &  24902   &   4.182  &  0.029  \\
1 & 200   &  23968   &   4.114  &  0.098  \\
1 & 300   &  23117   &   4.032  &  0.180  \\
1 & 400   &  21758   &   3.878  &  0.333  \\
2 & \phn 50    &  25256   &   4.008  &  0.011  \\
2 &100   &  24739   &   3.978  &  0.040  \\
2 &200   &  23653   &   3.898  &  0.120  \\
2 &300   &  22504   &   3.786  &  0.232  \\
2 &400   &  20575   &   3.568  &  0.449  \\
3 & \phn 50    &  25240   &   3.793  &  0.011  \\
3 & 100   &  24500   &   3.749  &  0.054  \\
3 & 200   &  23053   &   3.642  &  0.161  \\
3 & 300   &  21611   &   3.494  &  0.309  \\
3 & 400   &  18898   &   3.159  &  0.644  \\
4 & \phn 50    &  18466   &   4.236  &  0.018  \\
4 &100   &  18104   &   4.205  &  0.048  \\
4 &200   &  17235   &   4.119  &  0.135  \\
4 &300   &  16454   &   4.021  &  0.232  \\
4 &400   &  15185   &   3.841  &  0.412  \\
5 & \phn 50    &  18413   &   3.975  &  0.021  \\
5 & 100   &  17863   &   3.929  &  0.067  \\
5 & 200   &  16684   &   3.811  &  0.185  \\
5 & 300   &  15661   &   3.675  &  0.321  \\
5 & 400   &  13913   &   3.396  &  0.600  \\
6 & \phn 50    &  18361   &   3.773  &  0.024  \\
6 & 100   &  17643   &   3.713  &  0.085  \\
6 & 200   &  16656   &   3.604  &  0.194  \\
6 & 300   &  15165   &   3.403  &  0.395  \\
6 & 350   &  14178   &   3.233  &  0.565  \\
7 & \phn 50    &  15163   &   4.157  &  0.020  \\
7 & 100   &  14721   &   4.109  &  0.069  \\
7 & 200   &  14189   &   4.042  &  0.135  \\
7 & 300   &  13017   &   3.895  &  0.282  \\
7 & 350   &  12603   &   3.851  &  0.326  \\
8 & \phn 50    &  15110   &   3.952  &  0.025  \\
8 & 100   &  14588   &   3.896  &  0.081  \\
8 & 200   &  13827   &   3.796  &  0.181  \\
8 & 300   &  12709   &   3.656  &  0.321  \\
8 & 350   &  12053   &   3.561  &  0.416  \\
9 & \phn 50    &  15060   &   3.784  &  0.030  \\
9 & 100   &  14450   &   3.718  &  0.096  \\
9 & 200   &  13474   &   3.589  &  0.225  \\
9 & 300   &  12178   &   3.429  &  0.385  \\
\enddata
\end{deluxetable}

\clearpage
\begin{deluxetable}{lcccclccc}
\tablewidth{0pc}
\tablecaption{Parameters for Low Gravity Stars in the 
Middle and Low Mass Groups
\label{tab6}}
\tablehead{
\multispan{4}{\hfil Middle Mass Group ($\log g_{\rm polar} < 3.6$)\hfil} &
\multispan{5}{\hfil Low Mass Group ($\log g_{\rm polar} < 4.0$)\hfil} \\
\cline{1-4}
\cline{6-9}
\colhead{Cluster } &
\colhead{WEBDA\tablenotemark{a}  } &
\colhead{$T_{\rm eff}$ } &
\colhead{ } &
\colhead{} &
\colhead{Cluster } &
\colhead{WEBDA\tablenotemark{a}  } &
\colhead{$T_{\rm eff}$ } &
\colhead{ } \\
\colhead{Name } &
\colhead{Index } &
\colhead{(kK) } &
\colhead{$\log g_{\rm polar}$ } &
\colhead{} &
\colhead{Name } &
\colhead{Index } &
\colhead{(kK) } &
\colhead{$\log g_{\rm polar}$ }}
\startdata
IC2944\dotfill  &    \phn\phn 102  & 14.08 & 3.26 && IC1805\dotfill  &       \phn 1433 & 13.41 & 3.96 \\
NGC869\dotfill  &    \phn\phn 768  & 16.61 & 3.50 && NGC2384\dotfill &           10029 & 11.73 & 3.84 \\
NGC2362\dotfill &           10007  & 15.93 & 3.48 && NGC457\dotfill  &    \phn\phn 109 & 10.25 & 3.96 \\
NGC2422\dotfill & \phn\phn\phn 71  & 13.87 & 3.59 && NGC2244\dotfill & \phn\phn\phn 88 & 12.90 & 3.89 \\
NGC2422\dotfill & \phn\phn\phn 83  & 12.72 & 3.59 && NGC2244\dotfill &    \phn\phn 192 & 13.53 & 3.99 \\
NGC2467\dotfill & \phn\phn\phn 16  & 12.35 & 3.56 && NGC2422\dotfill & \phn\phn\phn 42 & 12.83 & 3.92 \\
NGC2467\dotfill & \phn\phn\phn 47  & 13.67 & 3.34 && NGC2467\dotfill &\phn\phn\phn\phn2& 11.88 & 3.94 \\
NGC2467\dotfill &           10017  & 15.11 & 3.50 && NGC2467\dotfill & \phn\phn\phn 33 & 12.81 & 3.89 \\
                &                  &       &      && NGC7160\dotfill &    \phn\phn 940 & 11.64 & 3.71 \\
\enddata
\tablenotetext{a}{Not a WEBDA number but assigned by us if $> 10000$.}
\end{deluxetable}

\clearpage  
\begin{deluxetable}{cccccccc}
\tabletypesize{\footnotesize}
\tablewidth{0pc}
\tablecaption{\ion{He}{1} Equivalent Widths from 
Solar Abundance LTE and non-LTE Models
\label{tab7}}
\tablehead{
\multispan{1}{ } &
\multispan{3}{\hfil Non-LTE (Lanz \& Hubeny)\hfil} &
\multispan{4}{\hfil LTE (Kurucz)\hfil} \\
\\
\cline{2-4}
\cline{6-8}
\colhead{$T_{\rm eff}$ } &
\colhead{\ion{He}{1} $\lambda 4026$ } &
\colhead{\ion{He}{1} $\lambda 4387$ } &
\colhead{\ion{He}{1} $\lambda 4471$ } &
\colhead{} &
\colhead{\ion{He}{1} $\lambda 4026$ } &
\colhead{\ion{He}{1} $\lambda 4387$ } &
\colhead{\ion{He}{1} $\lambda 4471$ } \\
\colhead{(K) } &
\colhead{(\AA) } &
\colhead{(\AA) } &
\colhead{(\AA) } &
\colhead{} &
\colhead{(\AA) } &
\colhead{(\AA) } &
\colhead{(\AA) }}
\startdata
\multispan{8}{\hfil $\log g = 3.0$\hfil} \\
       &         &           &          &&           &           &          \\
\tableline 
15000  &   0.772 &     0.427 &    0.743 &&     0.828 &     0.424 &    0.775 \\
17000  &   0.883 &     0.529 &    0.882 &&     0.903 &     0.498 &    0.882 \\
19000  &   0.831 &     0.526 &    0.849 &&     0.828 &     0.477 &    0.836 \\
21000  &   0.706 &     0.467 &    0.765 &&     0.676 &     0.399 &    0.704 \\
23000  &   0.605 &     0.407 &    0.661 &&     0.526 &     0.317 &    0.566 \\
25000  &   0.531 &     0.352 &    0.598 &&     0.378 &     0.235 &    0.429 \\
27000  &   0.474 &     0.279 &    0.537 &&     0.209 &     0.141 &    0.278 \\
\tableline \\
\multispan{8}{\hfil $\log g = 4.0$\hfil} \\ 
       &         &           &          &&           &           &          \\
\tableline 
15000  &   1.011 &     0.494 &    0.863 &&     1.121 &     0.495 &    0.941 \\
17000  &   1.338 &     0.731 &    1.201 &&     1.442 &     0.733 &    1.250 \\
19000  &   1.531 &     0.888 &    1.393 &&     1.579 &     0.857 &    1.411 \\
21000  &   1.498 &     0.895 &    1.385 &&     1.533 &     0.855 &    1.395 \\
23000  &   1.361 &     0.827 &    1.261 &&     1.384 &     0.779 &    1.268 \\
25000  &   1.211 &     0.745 &    1.147 &&     1.220 &     0.687 &    1.120 \\
27000  &   1.090 &     0.673 &    1.050 &&     1.076 &     0.606 &    0.994 \\
\enddata
\end{deluxetable}


\clearpage

\begin{thebibliography}{}
\bibitem[Abt et al.(2002)Abt, Levato, \& Grosso]{abt02}
         Abt, H. A, Levato, H., \& Grosso, M. 2002, \apj, 573, 359
\bibitem[Ardeberg \& Maurice(1977)]{ard77}
         Ardeberg, A., \& Maurice, E. 1977, \aaps, 28, 153
\bibitem[Arnal et al.(1988)]{arn88}
         Arnal, M., Morrell, N., Garcia, B., \& Levato, H. 1988, 
         \pasp, 100, 1076
\bibitem[Auer \& Mihalas(1972)]{aue72}
         Auer, L. H., \& Mihalas, D. 1972, \apjs, 24, 193
\bibitem[Auer \& Mihalas(1973)]{aue73}
         Auer, L. H., \& Mihalas, D. 1973, \apjs, 25, 433
\bibitem[Baume et al.(2003)]{bau03}
         Baume, G., Vazquez, R. A., Carraro, G., \& Feinstein, A. 2003, \aap, 402, 549
\bibitem[B\"ohm-Vitense(1981)]{boh81}
         B\"ohm-Vitense, E. 1981, \araa, 19, 295
\bibitem[Borra \& Landstreet(1979)]{bor79}
         Borra, E. F., \& Landstreet, J. D. 1979, \apj, 228, 809
\bibitem[Borra et al.(1983)Borra, Landstreet, \& Thompson]{bor83}
         Borra, E. F., Landstreet, J. D., \& Thompson, I. 1983, \apjs, 53, 151
\bibitem[Cardon de Lichtbuer(1982)]{car82}
         Cardon de Lichtbuer, P. 1982, Vatican Obs. Publ., 2, 1
\bibitem[Claret(1998)]{cla98}
         Claret, A. 1998, \aaps, 131, 395
\bibitem[Claret(2003)]{cla03}
         Claret, A. 2003, \aap, 406, 623
\bibitem[Claria et al.(2003)]{clar03}
         Claria, J. J., Lapasset, E., Piatti, A. E., \& Ahumada, A. V. 2003, \aap, 409, 541
\bibitem[Collins(1963)]{col63}
         Collins, G. W., II, 1963, \apj, 138, 1134
\bibitem[Collins \& Sonneborn(1977)]{col77}
         Collins, G. W., II, \& Sonneborn, G. H. 1977, \apjs, 34, 41
\bibitem[Collins et al.(1991)Collins, Truax, \& Cranmer]{col91}
         Collins, G. W., II, Truax, R. J., \& Cranmer, S. R. 1991, \apjs, 77, 541
\bibitem[Dachs \& Kaiser(1984)]{dac84}
         Dachs, J., \& Kaiser, D. 1984, \aaps, 58, 411
\bibitem[de Graeve(1983)]{deg83}
         de Graeve, E. 1983, Vatican Obs. Publ., 2, 31
\bibitem[Deutschman et al.(1976)Deutschman, Davis, \& Schild]{deu76}
         Deutschman, W. A., Davis, R. J., \& Schild, R. E. 1976, \apjs, 30, 97
\bibitem[Dombrowski \& Hagen-Thorn(1964)]{dom64}
         Dombrowski, W. A., \& Hagen-Thorn, W. A. 1964, Publ. Astron. Obs. Leningrad, 20, 75
\bibitem[Feinstein(1969)]{fei69}
         Feinstein, A. 1969, \mnras, 143, 273
\bibitem[Feinstein(1982)]{fei82}
         Feinstein, A. 1982, \aj, 87, 1012
\bibitem[Feinstein et al.(1973)Feinstein, Marraco, \& Muzzio]{fei73}
         Feinstein, A., Marraco, H. G., \& Muzzio, J. C. 1973, \aaps, 12, 331
\bibitem[Feinstein \& Vazquez(1989)]{fei89}
         Feinstein, A., \& Vazquez, R. A. 1989, \aaps, 77, 321
\bibitem[FitzGerald(1970)]{fit70}
         FitzGerald, M. P. 1970, \aap, 4, 234
\bibitem[Forbes et al.(1992)]{for92}
         Forbes, D., English, D., de Robertis, M. M., \& Dawson, P. C. 1992, \aj, 103, 916
\bibitem[Gies \& Lambert(1992)]{gie92}
         Gies, D. R., \& Lambert, D. L. 1991, \apj, 387, 673
\bibitem[Guetter(1964)]{gue64}
         Guetter, H. H. 1964, Publ. David Dunlap Obs., 2, 13
\bibitem[Gummersbach et al.(1998)]{gum98}
	 Gummersbach, C. A., Kaufer, A., Schaefer, D. R., Szeifert, T., \& Wolf, B.	
         1998, \aap, 338, 881
\bibitem[Heger \& Langer(2000)]{heg00}
         Heger, A., \& Langer, N. 2000, \apj, 544, 1016
\bibitem[Herbst \& Havlen(1977)]{her77}
         Herbst, W., \& Havlen, R. J. 1977, \aaps, 30, 279
\bibitem[Herbst \& Miller(1982)]{her82}
         Herbst, W., \& Miller, D. P. 1982, \aj, 87, 1478
\bibitem[Hill \& Lynas-Gray(1977)]{hil77}
         Hill, P. W., \& Lynas-Gray, A. E. 1977, \mnras, 180, 69
\bibitem[Hiltner(1956)]{hil56}
         Hiltner, W. A. 1956, \apjs, 2, 389
\bibitem[Hoag et al.(1961)]{hoa61}
         Hoag, A. A., Johnson, H. L., Iriarte, B., Mitchell, R. I., \& Hallam, K. L. 
         1961, Publ. USNO Second Ser., 17, 345
\bibitem[Huang \& Gies(2006)]{hua06}
         Huang, W., \& Gies, D. R. 2006, \apj, submitted (Paper I; astro-ph/0510450) 
\bibitem[Hubeny \& Lanz(1995)]{hub95}
         Hubeny, I., \& Lanz, T. 1995, \apj, 439, 875
\bibitem[Jackson et al.(2005)Jackson, MacGregor, \& Skumanich]{jac05}
	 Jackson, S., MacGregor, K. B., \& Skumanich, A. 2005, \apjs, 156, 245
\bibitem[Jaschek et al.(1968)]{jas68}
         Jaschek, C., Jaschek, M., Morgan, W. W., \& Slettebak, A.
         1968, \apj, 153, L87
\bibitem[Johnson(1962)]{joh62}
         Johnson, H. L. 1962, \apj, 136, 1135
\bibitem[Johnson \& Morgan(1953)]{joh53}
         Johnson, H. L., \& Morgan, W. W. 1953, \apj, 117, 313
\bibitem[Johnson \& Morgan(1955)]{joh55}
         Johnson, H. L., \& Morgan, W. W. 1955, \apj, 122, 429
\bibitem[Joshi \& Sagar(1983)]{jos83}
         Joshi, U. C., \& Sagar, R. 1983, J. RASC, 77, 40
\bibitem[Lanz \& Hubeny(2003)]{lan03}
         Lanz, T., \& Hubeny, I. 2003, \apjs, 146, 417
\bibitem[Lanz \& Hubeny(2005)]{lan05}
         Lanz, T., \& Hubeny, I. 2005, American Astronomical Society Meeting 207 Abstracts, \#182.21
\bibitem[Leone \& Manfre(1997)]{leo97}
         Leone, F., \& Manfre, M. 1997, \aap, 320, 257
\bibitem[Lod\'{e}n(1966)]{lod66}
         Lod\'{e}n, L. O. 1966, Arkiv f\"{o}r Astronomi, 4, 65
\bibitem[Loktin et al.(2000)Loktin, Gerasimenko, \& Malysheva]{lok01}
         Loktin, A.V., Gerasimenko, T.P., \& Malysheva, L.K.
         2001, Astron. Astrophys. Trans., 20, 607
\bibitem[Lynga(1959)]{lyn59}
         Lynga, G. 1959, Ark. Astron., 2, 379
\bibitem[Lyubimkov et al.(2004)Lyubimkov, Rostopchin, \& Lambert]{lyu04}
         Lyubimkov, L. S., Rostopchin, S. I., \& Lambert, D. L. 2004, \mnras, 351, 745
\bibitem[Massey \& Johnson(1993)]{mas93}
         Massey, P., \& Johnson, J. 1993, \aj, 105, 980
\bibitem[Massey et al.(1995)Massey, Johnson, \& DeGioia-Eastwood]{mas95}
         Massey, P., Johnson, K. E., \& DeGioia-Eastwood, K. 1995, \apj, 454, 151
\bibitem[Mathys(2004)]{mat04}
         Mathys, G.
         2004, in Stellar Rotation, Proc. IAU Symp. 215,
         ed. A. Maeder \& P. Eenens (San Francisco: ASP), 270
\bibitem[Mathys et al.(2002)]{mat02}
	 Mathys, G., Andrievsky, S. M., Barbuy, B., Cunha, K., \& Korotin, S. A.
         2002, \aap, 387, 890
\bibitem[McAlister et al.(2005)]{mca05}
         McAlister, H. A., et al. 2005, \apj, 608, 439
\bibitem[Mermilliod \& Paunzen(2003)]{mer03}
         Mermilliod, J.-C., \& Paunzen, E. 2003, \aap, 410, 511
\bibitem[Meynet \& Maeder(1997)]{mey97}
         Meynet, G., \& Maeder, A. 1997, \aap, 321, 465
\bibitem[Meynet \& Maeder(2000)]{mey00}
         Meynet, G., \& Maeder, A. 2000, \aap, 361, 101
\bibitem[Moffat \& Vogt(1974)]{mof74}
         Moffat, A. F. J., \& Vogt, N. 1974, Veroeff. Astron. Inst. Bochum, 2, 1
\bibitem[Ogura \& Ishida(1981)]{ogu81}
         Ogura, K., \& Ishida, K. 1981, Publ. Astron. Soc. Japan, 33, 149
\bibitem[Pedreros(1984)]{ped84}
         Pedreros, M. H. 1984, Ph.D. thesis (Univ. Toronto, David Dunlap Obs.)
\bibitem[Perez \& Westerlund(1987)]{per87}
         Perez, M. R., \& Westerlund, B. E. 1987, \pasp, 99, 1050
\bibitem[Perry(1973)]{per73}
         Perry, C. L. 1973, in 
         Spectral Classification and Multicolour Photometry, IAU Symp. 50,
         ed. C. Fehrenbach \& B. E. Westerlund (Dordrecht: Reidel), 192
\bibitem[Perry et al.(1976)]{per76}
         Perry, C. L., Franklin, C. B., Landolt, A. U., \& Crawford, D. L. 1976, \aj, 81, 632
\bibitem[Pesch(1959)]{pes59}
         Pesch, P. 1959, \apj, 130, 764
\bibitem[Prisinzano et al.(2003)]{pri03}
         Prisinzano, L., Micela, G., Sciortino, S., \& Favata, F. 2003, \aap, 404, 927
\bibitem[Purgathofer(1964)]{pur64}
         Purgathofer, A. 1964, Ann. Univ. Sternw. Wien, 26, 2
\bibitem[Reimann \& Pfau(1987)]{rei87}
         Reimann, H.-G., \& Pfau, W. 1987, Astron. Nach., 308, 111
\bibitem[Sackmann \& Anand(1970)]{sac70}
	 Sackmann, I.-J., \& Anand, S. P. S. 1970, \apj, 162, 105
\bibitem[Sanner et al.(2001)]{san01}
         Sanner, J., Brunzendorf, J., Will, J.-M., \& Geffert, M. 2001, \aap, 369, 511
\bibitem[Schaller et al.(1992)]{sch92}
         Schaller, G., Schaerer, D., Meynet, G., \& Maeder, A.   
         1992, \aaps, 96, 269
\bibitem[Schild(1965)]{sch65}
         Schild, R. E. 1965, \apj, 142, 979
\bibitem[Shore et al.(2004)]{sho04}
         Shore, S. N., Bohlender, D. A., Bolton, C. T., North, P., \& Hill, G. M. 2004, \aap, 421, 203
\bibitem[Slesnick et al.(2002)Slesnick, Hillenbrand, \& Massey]{sle02}
         Slesnick, C. L., Hillenbrand, L. A., \& Massey, P. 2002, \apj, 576, 880
\bibitem[Tapia et al.(1984)]{tap84}
         Tapia, M., Roth, M., Costero, R., \& Navarro, S. 1984, Revista Mex. Astron. Astrofis., 9, 65
\bibitem[Tapia et al.(2003)]{tap03}
         Tapia, M., Roth, M., Vazquez, R. A., \& Feinstein, A. 2003, \mnras, 339, 44
\bibitem[Thackeray \& Wesselink(1965)]{tha65}
         Thackeray, A. D, \& Wesselink, A. J. 1965, \mnras, 131, 121
\bibitem[Turner(1976)]{tur76}
         Turner, D. G. 1976, \apj, 210, 65
\bibitem[Turner et al.(1980)]{tur80}
         Turner, D. G., Grieve, G. R., Herbst, W., \& Harris, W. E. 1980, \aj, 85, 1193
\bibitem[Underhill et al.(1979)]{und79}
         Underhill, A. B., Divan, L., Prevot-Burnichon, M.-L., \& Doazan, V. 1979, \mnras, 189, 601
\bibitem[Vazquez \& Feinstein(1991)]{vaz91}
         Vazquez, R. A., \& Feinstein, A. 1991, \aaps, 92, 863
\bibitem[Vogt \& Moffat(1972)]{vog72}
         Vogt, N., \& Moffat, A. F. J. 1972, \aaps, 7, 133
\bibitem[Wade et al.(1997)]{wad97}
         Wade, G. A., Bohlender, D. A., Brown, D. N., Elkin, V. G., Landstreet, J. D., \& Romanyuk, I. I.
         1997, \aap, 320, 172
\bibitem[von Zeipel(1924)]{von24}
         von Zeipel, H. 1924, \mnras, 84, 665
\end{thebibliography}
\end{document}